\documentclass[final,5p,times,twocolumn]{elsarticle}

\usepackage[colorlinks=true, urlcolor=MidnightBlue, linkcolor=MidnightBlue, 
citecolor=Mahogany, pdfborder={0 0 0}]{hyperref}
\usepackage{amssymb}
\usepackage{lipsum}
\usepackage{siunitx}
\usepackage{booktabs}
\usepackage{multirow}
\usepackage{amsthm}
\usepackage{amsmath}
\usepackage{verbatim}
\usepackage{tabularray}
\usepackage{makecell}
\usepackage{lineno}

\newcommand{\angstrom}{\r{A}}
\usepackage{xcolor}

\journal{Acta Materialia}

\begin{document}

\begin{frontmatter}


\title{A Combined DFT and MD Study on Interface Stability in Ferrite-Cementite Systems}

\author[1]{Pablo Canca}
\ead{pcanca@ugr.es}
\author[2]{Chu-Chun Fu}
\author[3]{Christophe J. Ortiz}
\author[1,4]{Blanca Biel\corref{cor1}}
\ead{biel@ugr.es}
\cortext[cor1]{Corresponding author}
\affiliation[1]{organization={Dpto. Física Atómica, Molecular y Nuclear. Facultad de Ciencias, Campus de Fuente Nueva, Universidad de Granada},
            city={Granada},
            postcode={18071}, 
            country={Spain}}

\affiliation[2]{organization={Université Paris-Saclay, CEA, Service de recherche en Corrosion et Comportement des Matériaux, SRMP}, 
            city={Gif-sur-Yvette},
            postcode={F-91191}, 
            country={France}}

\affiliation[3]{organization={Laboratorio Nacional de Fusión por Confinamiento Magnético - CIEMAT},
            city={Madrid},
            postcode={28040}, 
            country={Spain}}

\affiliation[4]{organization={Instituto Carlos I de Física Teórica y Computacional, Facultad de Ciencias, Campus de Fuente Nueva, Universidad de Granada},
            city={Granada},
            postcode={18071}, 
            country={Spain}}

\begin{abstract}

{
Understanding the atomic structure and energetic stability of ferrite–cementite interfaces is essential for optimizing the mechanical performance of steels, especially under extreme conditions such as those encountered in nuclear fusion environments. In this work, we combine Classical Molecular Dynamics (MD) and Density Functional Theory (DFT) to systematically investigate the stability of ferrite–cementite interfaces within the Bagaryatskii Orientation Relationship. Three interface orientations and several cementite terminations are considered to identify the most stable configurations.

MD simulations reveal that the $(010)_{\theta}||(11\bar{2})_{\alpha}$ and $(001)_{\theta}||(1\bar{1}0)_{\alpha}$ orientations are energetically favourable for selected terminations, and these predictions are validated and refined by subsequent DFT calculations. A key result of our study is the destabilizing effect of interfacial carbon atoms, which increase the interface energy and decrease the Griffith energy, indicating a reduced resistance to fracture. This finding contrasts with earlier reports suggesting a stabilizing role for carbon.

Our analysis of the electronic structure shows that Fe–C bonding at the interface perturbs the metallic environment of interfacial Fe atoms. This bonding response explains the observed variations in magnetic moment and helps rationalize the trends in interface energy. We also establish correlations between interface energy, magnetic perturbation, and a bond-based descriptor quantifying new and broken bonds. These insights provide a physically grounded, predictive framework for the design and optimization of ferrite–cementite interfaces in advanced steels.
}
\end{abstract}




\end{frontmatter}


\section{Introduction}
\label{introduction}

Steels have become one of the most widely used materials in construction, technology and engineering due to their outstanding mechanical properties. However, in extreme conditions such as those inside fusion reactors, novel steels with stronger resistance to irradiation are needed. EUROFER97, a reduced activation ferritic/martensitic (RAFM) steel, is a key candidate for the DEMO fusion reactor's critical components \cite{rieth2003eurofer}. It is indeed expected to endure severe neutron irradiation that will generate different types of bulk defects, which will in turn affect its macroscopic properties \cite{gaganidze2013assessment, lucas1993evolution, field2017mechanical}. Understanding steel behaviour at the atomic level is thus crucial to anticipate these effects.

Carbon complexes have a significant effect on the evolution of steel structures under irradiation and are partly responsible for damage accumulation, as suggested by different studies \cite{ortiz2009influence, domain2004ab, candela2018interaction, xu2013solving}. Despite considerable research, many aspects of the mechanisms involved in the arrangement of C atoms in a bcc-Fe lattice remain unclear. For example, several mechanisms explaining the redistribution of C atoms dissolved as interstitials in Fe-fcc austenite into the formation of cementite nanocrystals in ferrite are still not fully understood \cite{prester2022migration, zhang2015structural, buggenhoudt2021predicting}. Additionally, there is limited information regarding the segregation of C atoms into or out of the interface formed between cementite and ferrite.

Over the past decades, three main orientation relationships (ORs) between the cementite precipitate and the ferrite host lattice have been reported in the literature: the Isaichev \cite{isaichev1947orientation}, the Bagaryatskii \cite{bagaryatskii1950probable}, and the Pitsch-Petch \cite{petch1953orientation, pitsch1962orientierungszusammenhang}. Variations of these interfaces, referred to as Near Bagaryatskii and Near Pitsch-Petch \cite{kim2020effect}, have also been documented in recent years. Experimental validation of these results is challenging due to the complexity of determining the OR between the precipitate —cementite in this case— and the ferrite lattice. Therefore, accurate computational methods play an essential role in understanding the interface configuration.

Although the Bagaryatskii OR has been extensively studied using \textit{ab initio} techniques \cite{guziewski2016atomistic, wang2018interface, RUDA2009550}, several open questions remain. First, discrepancies between methodologies and results exist in the reported interface energy for this OR. Second, while it is well known that classic Molecular Dynamics (MD) methods typically fail to accurately reproduce structural and energy properties compared to Density Functional Theory (DFT), none of these studies calculate the interface energy $E_{\text{int}}$ using DFT methods. To our knowledge, only Zhang \textit{et al.} mention having calculated $E_{\text{int}}$ for the Bagaryatskii OR using DFT \cite{zhang2015structural}, although no details are provided regarding the calculation method. Finally, the treatment of non-stoichiometric terminations of cementite is not clearly defined in the literature. In this work, we offer new insights into key aspects of this interface by applying a more rigorous approach and a more precise methodology to calculate the interface energy using DFT.

The Bagaryatskii OR follows the relationship:

\begin{align*}
[100]_{\theta}||[111]_{\alpha}\\
(010)_{\theta}||(11\bar{2})_{\alpha}\\
[001]_{\theta}||[1\bar{1}0]_{\alpha}
\end{align*}

where $\theta$ corresponds to the cementite lattice and $\alpha$ to the host lattice, ferrite.

\begin{figure}[ht]
    \centering
    \includegraphics[width=0.45\textwidth]{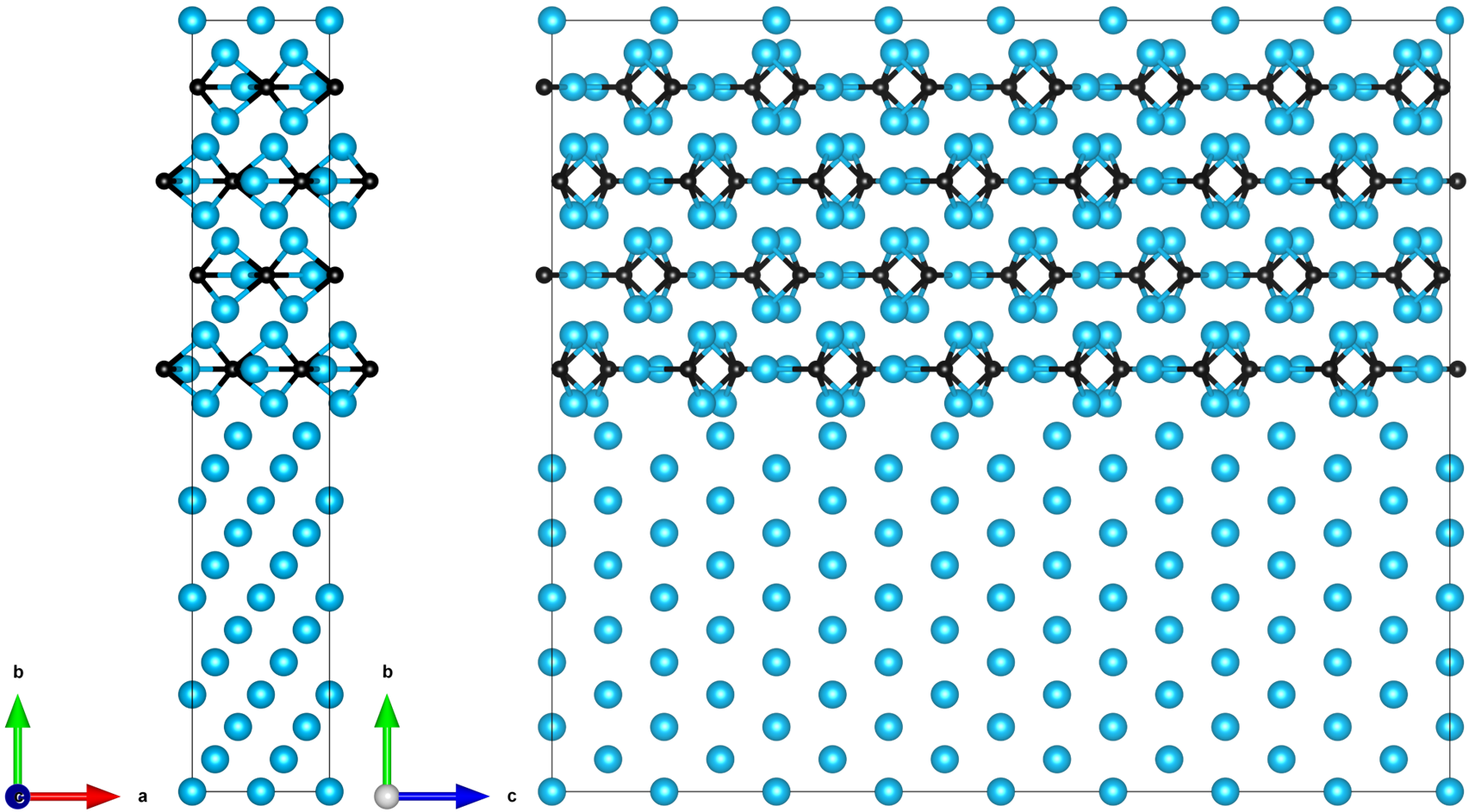}
    \caption{Ferrite/cementite interface in the Bagaryatskii OR for the interface plane $(010)_{\theta}||(11\bar{2})_{\alpha}$ with stoichiometric cementite. The larger blue spheres represent iron, while the smaller black spheres represent carbon. The direction $a$ corresponds to $[100]_{\theta}||[111]_{\alpha}$, and $c$ to $[001]_{\theta}||[1\bar{1}0]_{\alpha}$.}
    \label{fig:bag_OR}
\end{figure}

In this geometry, the interface plane is $(010)_{\theta}$ for cementite, which is symmetrical to the $(11\bar{2})_{\alpha}$ plane of Fe. However, in the case where a cementite nanocrystal is surrounded by a ferrite lattice, alternative configurations may arise. In these cases, the interface plane might be oriented differently, potentially as $(100)_{\theta}||(111)_{\alpha}$ or $(001)_{\theta}||(1\bar{1}0)_{\alpha}$ \cite{wang2018interface}. To date, these orientations have not been investigated using DFT-based computational studies, leaving open the possibility that the commonly observed $(010)_{\theta}||(11\bar{2})_{\alpha}$ interface plane may not be the most energetically stable. This study will consider these three possible configurations. 

Moreover, the configuration of the interface could also be influenced by the termination planes of both crystals, a factor that has been largely overlooked in previous computational studies. While the termination plane does not significantly affect the structure of ferrite due to its high symmetry, in cementite, it plays a crucial role in determining the interfacial energy and must be carefully considered.

Several classical MD potentials (such as the Embedded Atom Method (EAM) of Ruda \cite{RUDA2009550}, the EAM of Hepburn \cite{hepburn2008metallic}, the Tersoff method of Henriksson \cite{Tersoff_PhysRevB.79.144107}, and the Modified Embedded Atom Method (MEAM) of Liyanage \cite{liyanage2014structural}), along with various approaches, applied strains (to either the ferrite or the cementite lattice), and cell sizes, have been used in previous studies to investigate these interfaces, leading to irreconcilable conclusions \cite{guziewski2016atomistic, wang2018interface, RUDA2009550}. For example, $E_{int}$ ranges from 0.1 J/m², as predicted by Wang \textit{et al.} \cite{wang2018interface}, to 1.23 J/m², as predicted by Guziewski \textit{et al.} \cite{guziewski2016atomistic}, both using EAM potentials, for the stoichiometric $(010)_{\theta}||(11\bar{2})_{\alpha}$ interface plane.

Therefore, a more accurate approach is to use DFT. However, the large number of potential structures to investigate, combined with the significant interface sizes required to minimise strain from the cementite-ferrite mismatch, presents computational challenges that have not been fully addressed.

In this work, we use MD as a preliminary tool to optimise the search for energetically stable interfaces within the Bagaryatskii OR. We first use classic MD to explore different orientations, termination planes, relative positions, and other micro degrees of freedom within our systems. We then focus on those structures with sufficiently low interface energy---indicative of their prevalence in nature \cite{wulff1901xxv}---for further investigation using DFT, which allows not only to obtain more accurate results, but also to extract information about the electronic structure of the interfaces.

The stability of surfaces and interfaces explored in this work is assessed by analysing the number of broken and newly formed bonds, weighted by their bond strength \cite{barbe2018fracture, sonderegger2010interfacial}. In this paper, we implement a predictive technique for estimating the interfacial energy of $\alpha-\theta$ interfaces, which we subsequently verify through bond order calculations (which is proportional to bond strength) using the Density Derived Electrostatic and Chemical (DDEC6) approach \cite{manz2016introducing}. Finally, we also study Griffith energy as a first approximation to interface fracture.

The structure of the paper is as follows: in section 2, we will address the methodology followed for all calculations using both MD, DFT and how the bond analysis is carried out. Section 3 will detail the results obtained using MD for a big, low strain interface and for a small, high strain, DFT adapted interface, as well as the different surfaces. The influence of the added strain is assessed. Finally, section 4 shows the results obtained with DFT, bond analysis and its correlation with our results, the calculation of Griffith energy and the electronic structure of the system.

\section{Methodology}

\subsection{Energetic models and simulation set-up}

Interface systems typically require a significant number of atoms to minimise strains arising from lattice constant mismatches between the two crystals---in this case, cementite and ferrite---leading to misfits.

However, using techniques suitable for such large structures often compromises results accuracy. To address this challenge, we propose a method that integrates MD and DFT. Initially, we investigate interfaces with minimal strain using classic MD, which allows the treatment of a large number of atoms. This initial screening can already help to discard several orientations and terminations due to their high interface energies. After this initial characterisation, we construct a smaller interface---thus with higher strain---to be analysed with DFT.

In order to account for the impact of the added strain due to mismatch, this interface undergoes a preliminary investigation using MD.  This is required to assess other factors that influence $E_{int}$ regardless of the method employed, such as strain or the relative position of the crystals forming the interface, which, due to the high number of possibilities, cannot be studied using DFT alone. Subsequently, a similarly sized structure is analysed using DFT, which provides the target outcomes. This combined method allows to calculate high accuracy, DFT results, which would be otherwise impossible to obtain, thanks to MD.

\subsubsection{Molecular Dynamics Calculations} \label{sec:Molecular_Dynamics_Methods}

MD simulations were performed using LAMMPS (Large-scale Atomic/Molecular Massively Parallel Simulator) \cite{LAMMPS}. The conjugate-gradient algorithm was used to minimise the system's energy, with stopping criteria set to $10^{-10}$ for energy and $10^{-6}$ for force. Periodic boundary conditions were applied in all three directions.

Several potentials were tested to verify their accuracy in reproducing the ferrite/cementite system, focusing on lattice constants and the formation energy of cementite. Among these potentials, we tested the EAM by Hepburn \cite{hepburn2008metallic}, the Tersoff by Henriksson \cite{Tersoff_PhysRevB.79.144107} and the MEAM by Liyanage \cite{liyanage2014structural}. Of these, the MEAM provided results closest to DFT, which are summarised in Table \ref{tab:constants_potentials} (see the Supplementary Material for the results with the other potentials).

\begin{table}[ht]
    \centering
    \setlength{\tabcolsep}{2pt}
    \begin{tabular}{ccccc}
        \toprule
        & MD & DFT & Experimental & $\varepsilon_\text{MD/DFT}$ (\%)\\
        \midrule
        \midrule
        $a_{Fe}$ (\angstrom) & 2.851 & 2.833 \cite{haas2009calculation} & 2.853 \cite{haas2009calculation} & 0.6 \\
        $a_{Cem}$ (\angstrom) & 5.088 & 5.036 \cite{buggenhoudt2021predicting} & 5.081  \cite{wood2004thermal} & 1.0 \\
        $b_{Cem}$ (\angstrom) & 6.669 & 6.721 \cite{buggenhoudt2021predicting} & 6.753 \cite{wood2004thermal} & 0.8 \\
        $c_{Cem}$ (\angstrom) & 4.470 & 4.479 \cite{buggenhoudt2021predicting} & 4.515 \cite{wood2004thermal} & 0.2 \\
        $E^{f}_{Cem}$ (eV/f.u.) & 0.225 & 0.204 \cite{buggenhoudt2021predicting} & $0.165\pm0.05$ \cite{wood2004thermal} & 10.3 \\
        \bottomrule
    \end{tabular}
    \caption{Lattice constants of $\alpha$-Fe and cementite, as well as the formation energy of cementite, obtained with MD using the MEAM potential by Liyanage \cite{liyanage2014structural}. Results are compared with DFT results using GGA-PBE and experimental data from the literature. The relative error $\varepsilon$ between our MD results and the reference DFT values is also calculated.}
    \label{tab:constants_potentials}
\end{table}

As observed in Table \ref{tab:constants_potentials}, the lattice constants obtained with the MEAM potential accurately match those computed with DFT. This allows to build interfaces of similar size and strain. The formation energy of cementite, which is 10\% higher with the MEAM potential than with DFT, lies within an acceptable difference.

In a preliminary study, the ferrite/cementite interfaces were modelled using two distinct approaches: the bi-crystal approximation, which generates two interfaces, and slabs, where one interface is embedded in a vacuum, thereby introducing two additional surfaces. After conducting various tests, we chose the bi-crystal approximation to avoid the additional error introduced by surface energies. However, slabs are used for the study of surfaces and to understand the required size of the supercell that reproduces bulk conditions.

The Bagaryatskii OR can be classified as a semi-coherent interface due to its misfit of 3\%, 4\%, and 10\% in the $[100]_{\theta}||[111]_{\alpha}$, $[010]_{\theta}||[11\bar{2}]_{\alpha}$, and $[001]_{\theta}||[1\bar{1}0]_{\alpha}$ directions, respectively. Although this misfit is significant, it does not reach the levels characteristic of an incoherent interface. To mitigate excessively large misfits, which would result in unrealistically high strains, we employed sufficiently large supercells. All studied systems were built large enough to ensure strains below 0.22\% in the directions parallel to the interface.

Similarly, it is crucial for the interfaces to be large enough in the direction perpendicular to the interface plane to accurately replicate bulk conditions and prevent interactions between interfaces. To determine the necessary size before constructing the interfaces, we conducted tests on cementite and ferrite slabs. These tests involved computing the surface energy of each crystal at different orientations while progressively increasing the crystal's thickness perpendicular to the surface. We observed that the surface energy reached a constant value when the lattice was sufficiently large to prevent interaction between surfaces, typically occurring at approximately 24 \angstrom\, for cementite and 14 \angstrom\, for ferrite.

To quantify the influence of strain on our results, which cannot be avoided due to periodic boundary conditions and the mismatch between the lattice constants of cementite and ferrite, we investigate the reduced-size, high-strain interfaces using MD as an intermediate step before transitioning to DFT. For these smaller interfaces, all atomic positions near the interface were allowed to relax, except in a specified region where atom movement was restricted to replicate bulk conditions. The lattice vector perpendicular to the interface plane was also allowed to relax.

The effect of the relative position between ferrite and cementite was studied for all interfaces. In these calculations, all atoms were fixed, and the lattice vector perpendicular to the interface was allowed to relax.

\subsubsection{Density Functional Theory} \label{sec:DFT_methods}

The interfaces with the lowest energy, as determined by the MD results, were subsequently analysed using DFT.

DFT calculations were performed using the Vienna Ab Initio Simulation Package (VASP) \cite{kresse1996efficient} with the Projector-Augmented-Wave (PAW) method \cite{marsman_2016, blochl1994projector}. We employed the GGA exchange-correlation functional with the Perdew-Burke-Ernzerhof (PBE) scheme \cite{paier2005perdew}. Calculations were spin-polarised within the collinear approximation. The plane-wave energy cutoff was set to 500 eV for all calculations. The valence electrons for Fe are \textit{3d} and \textit{4s}, while for C, they are \textit{2s} and \textit{2p}. Periodic boundary conditions were applied in all three directions.

$\Gamma$-centered $k$-point meshes were generated using the Monkhorst-Pack scheme \cite{monkhorst1976special}. A grid of 9x9x9 k-points was used for characterising one unit cell of cementite. A similar grid was employed for a unit cell of rotated ferrite, which requires 12 atoms to accurately represent the orientation in space.

Convergence tests yielded a lattice constant of $a_{\text{Fe}} = 2.832$ Å for ferrite. For cementite, we obtained $a_{\text{cem}} = 5.032$ Å, $b_{\text{cem}} = 6.721$ Å, and $c_{\text{cem}} = 4.488$ Å, which are in close agreement with values reported in the literature  (see Table \ref{tab:constants_potentials}).

The convergence cutoff for the electronic self-consistency loop was set to $10^{-6}$ eV. For structural optimisations, the cutoff was 0.02 eV/Å. Structural optimisations were conducted at constant volume for slab calculations. For interfaces modelled using the bi-crystal approximation, the cell's lattice vector perpendicular to the interface was allowed to relax.

All systems were constructed to be sufficiently large, ensuring that the magnetic moments of atoms distant from the interface deviated by no more than 2\% from bulk values in ferrite and 3.5\% in cementite. In the bulk structure, ferrite Fe atoms exhibit a magnetic moment of 2.2 $\mu_B$, while cementite Fe atoms exhibit values of 1.96 $\mu_B$ and 1.89 $\mu_B$ at positions I and II, respectively. Carbon atoms in cementite have a magnetic moment of -0.12 $\mu_B$ \cite{buggenhoudt2021predicting}. Magnetic moments are calculated for strained structures as well and compared with the interface values.

\subsection{Calculated properties}
\subsubsection{Surface-formation energy}
We begin our study of the Bagaryatskii OR by analysing the surface-formation energy density ($\gamma_{\text{S}}$), commonly referred to as surface energy, of the studied systems.

For a stoichiometric system, the surface energy $\gamma_\text{S}$ is given by:

\begin{equation}
\gamma_\text{S}=\frac{1}{2A}\left(E_{\text{slab}}-E_{\text{bulk}}\right)=\frac{1}{2A}\left(E_{\text{slab}}-\sum_{i=0}^{n}N_{i}\mu_{i}\right)\label{eq:gamma_surface}
\end{equation}

Here, $E_{\text{slab}}$ is the computed energy of the crystal's slab, $N_{i}$ represents the number of atoms of species $i$, and $\mu_{i}$ is the chemical potential of species $i$.

The surface energies of stoichiometric ferrite and cementite obtained using MD with the MEAM potential by Liyanage \cite{liyanage2014structural} and DFT are presented in Table \ref{tab:Surface-energy} for different orientations.

\begin{table}[ht]
    \centering
        \begin{tabular}{cccc}
            \toprule
            $\gamma_{S}\,(\text{J/m}^{2})$ & MD & DFT & Reference (method) \\
            \midrule
            \midrule
            $\left(001\right)_{\alpha}$ & 2.10 & 2.50 & 2.5 \cite{tran2016surface} (DFT) \\
            \midrule
            $\left(111\right)_{\alpha}$ & 2.32 & 2.71 & 2.73 \cite{tran2016surface} (DFT)\\
            \midrule
            $(11\bar{2})_{\alpha}$ & 2.24 & 2.58 & 2.61 \cite{tran2016surface} (DFT) \\
            \midrule
            $(1\bar{1}0)_{\alpha}$ & 1.99 & 2.44 & 2.45 \cite{tran2016surface} (DFT)\\
            \midrule
            $\left(100\right)_{\theta}$ & 2.01 & 2.64 & 2.34 \cite{RUDA2009550} (MD); 2.64 \cite{broos2020quantum} (DFT) \\
            \midrule
            $\left(010\right)_{\theta}$ & 1.80 & 2.46 & 2.00 \cite{RUDA2009550} (MD); 2.46 \cite{broos2020quantum} (DFT) \\
            \midrule
            $\left(001\right)_{\theta}$ & 2.05 & 2.20 & 1.96 \cite{RUDA2009550} (MD); 2.21 \cite{broos2020quantum} (DFT) \\
            \bottomrule
        \end{tabular}
    \caption{Surface energy $\gamma_{S}$ of ferrite and cementite for different orientations. The results are given for unstrained cementite and are computed using the MEAM potential by Liyanage \cite{liyanage2014structural} for MD calculations and GGA-PBE for DFT calculations. The values are compared with literature references: \cite{tran2016surface} (GGA-PBE), \cite{RUDA2009550} (EAM), and \cite{chiou2003structure} (GGA-PW91).}
    \label{tab:Surface-energy}
\end{table}
 
Our DFT results for ferrite accurately match those reported in the literature \cite{RUDA2009550, tran2016surface, chiou2003structure}. The MD results differ due to the different nature of both methods, yet they correctly predict the relative stability of each surface. Specifically, the $(1\bar{1}0)_{\alpha}$ plane has the lowest surface energy $\gamma_{S}$, while the $\left(111\right)_{\alpha}$ plane has the highest.

However, discrepancies between MD and DFT become evident when examining the results for cementite. In our MD simulations using the MEAM potential, the $\left(010\right)_{\theta}$ surface shows the lowest $\gamma_{\text{S}}$, whereas DFT calculations and the EAM potential by Ruda {\it et al.} \cite{RUDA2009550} favour the $\left(001\right)_{\theta}$ surface energetically. This suggests that the EAM potential might better describe cementite surfaces compared to the MEAM potential by Liyanage \cite{liyanage2014structural}. Nevertheless, the interest of using MD is to reduce the number of DFT calculations, which provide the accurate results. In this way, the MEAM potential performs better at predicting lattice constants and formation energy, which are critical for accurately translating MD-generated structures into DFT simulations. The EAM potential, while providing different surface energy results, does not align as well with these key parameters, and especially with the formation energy of cementite, which yields results ten times higher than the experimental ones. Furthermore, the presence of carbon atoms requires potentials that consider bond orientation, and so we choose to continue using the MEAM in our MD calculations.

Thus far, our discussion has primarily focused on stoichiometric cementite. However, it is crucial to recognise that the structure and chemical properties of crystal surfaces can vary significantly depending on the chemical termination. While the structure of bcc Fe remains consistent across its terminating planes due to its high symmetry, cementite exhibits significant structural changes depending on the terminating plane.

Different terminating planes are obtained by removing an atomic layer from the surface of the cementite crystal, disrupting the stoichiometry of the system. This disruption complicates the calculation of bulk energy by means of  \textit{ab initio} methods because periodic boundary conditions would introduce unwanted interfaces.

To address this issue, we employ the method developed by Rapcewicz \textit{et al}. \cite{rapcewicz1998consistent}, which has been used by other researchers \cite{barbe2018fracture, abavare2014surface, siegel2002adhesion} in similar cases. This method provides a consistent approach to account for the changes in surface structure and chemical termination, allowing for accurate energy calculations in the presence of periodic boundary conditions.

The chemical potential of cementite's unit formula can be expressed as:

\begin{equation}
\mu_{\text{Fe}_{3}\text{C}}=3\mu_{\text{Fe}}+\mu_{\text{C}}
\end{equation}

where $\mu_{\text{Fe}}$ and $\mu_{\text{C}}$ are the chemical potentials of Fe and C in cementite, respectively. Since cementite is thermodynamically stable \cite{bhadeshia2020cementite}, the chemical potentials of Fe and C in the carbide must be lower than or equal to those of the pure systems:

\begin{equation}
\mu_{\text{Fe}}\leq\mu_{\text{Fe}}^{P};\,\mu_{\text{C}}\leq\mu_{\text{C}}^{P}
\end{equation}

where $\mu_{\text{Fe}}^{P}$ and $\mu_{\text{C}}^{P}$ are the chemical potentials of Fe and C in their pure forms (i.e., Fe-bcc and graphite). In our case, where cementite is relatively small compared to the surrounding Fe-bcc, we approximate $\mu_{\text{Fe}}\approx\mu_{\text{Fe}}^{P}$. 

Thus,

     \begin{align}
        \sum_{i=0}^{n}N_{i}\mu_{i}
        & = N_{\text{Fe}}\mu_{\text{Fe}}+N_{\text{C}}\mu_{\text{C}}=N_{\text{Fe}_{3}\text{C}}\mu_{\text{Fe}_{3}\text{C}}+N_{\text{Fe}}^{\text{ex}}\mu_{\text{Fe}}+N_{\text{C}}^{\text{ex}}\mu_{\text{C}}= \notag \\
        & = N_{\text{Fe}_{3}\text{C}}\mu_{\text{Fe}_{3}\text{C}}+N_{\text{Fe}}^{\text{ex}}\mu_{\text{Fe}}^{P}+N_{\text{C}}^{\text{ex}}\left(\mu_{\text{Fe}_{3}\text{C}}-3\mu_{\text{Fe}}^{P}\right)
     \end{align}

where $N_{i}^{\text{ex}}$ is the number of extra or lacking atoms of species $i$. Therefore, the surface energy for non-stoichiometric cementite, Eq. (\ref{eq:gamma_surface}), can be expressed as:

\begin{equation}
    \gamma_{\text{S}}^{\text{Cem}}=\frac{1}{2A}\left(E_{\text{slab}}-\mu_{\text{Fe}_{3}\text{C}}\left(N_{\text{Fe}_{3}\text{C}}+N_{\text{C}}^{\text{ex}}\right)-\mu_{\text{Fe}}^{P}\left(N_{\text{Fe}}^{\text{ex}}-3N_{\text{C}}^{\text{ex}}\right)\right)
    \label{eq:gamma_non_st}
\end{equation}

And so the bulk energy of a non-stoichiometric cementite crystal will be given by

\begin{equation}
E_{\text{bulk}}=\mu_{\text{Fe}_{3}\text{C}}\left(N_{\text{Fe}_{3}\text{C}}+N_{\text{C}}^{\text{ex}}\right)+\mu_{\text{Fe}}^{P}\left(N_{\text{Fe}}^{\text{ex}}-3N_{\text{C}}^{\text{ex}}\right)
\end{equation}
 
\subsubsection{Interface-formation energy} \label{sec:Eint}

In this study, we adopt the bi-crystal approach to model the interface between ferrite and cementite. This method helps to minimise the total error in the interface energy calculation by avoiding the need to separately account for surface energy, which can be highly sensitive to calculation procedures.

The interface energy density ($E_\text{int}$), often referred to simply as the interface energy, for the case of a homogeneous bi-crystal interface is given by:

\begin{equation}
    E_{\text{int}}=\frac{E_{\text{total}}-E_{\alpha}^{\text{bulk}}-E_{\theta}^{\text{bulk}}}{2S}\,\left(\text{J}\text{m}^{-2}\right)
\label{eq:biint}
\end{equation}

where $E_{\text{int}}$ represents the interface energy density, $E_{\text{total}}$ is the total energy of the interface system, $E_{\alpha}^{\text{bulk}}$ is the energy of bulk ferrite, $E_{\theta}^{\text{bulk}}$ is the energy of bulk cementite with the appropriate strain, and $S$ is the surface area of the interface. This partially compensates for the error introduced by strain.

\subsubsection{Bond Analysis}

We use the Chargemol code \cite{limas2018introducing} to obtain the bond orders (BO) of the bulk structures. VASP calculations reconstruct the all-electron charge density on a FFT grid with sufficient accuracy, which Chargemol then uses for the BO analysis. This allows us to obtain a magnitude proportional to bond strength between all atoms in the lattice. 

We denote as $\Omega_\text{BBS}$ a measure of the stability of surfaces normalised by the number of atoms at the surface, which is given by

\begin{equation}\label{eq:bb_surf}
    \Omega_\text{SBB} = \frac{1}{N_\text{S}}\sum_i^n N_{\text{b}i} \times Q_{i}
\end{equation}

where $N_\text{S}$ is the number of atoms at the surface, $N_{\text{b}i}$ is the number of broken bonds for each type of bod $i$ and $Q_{i}$ is a measure of the bond strength of each type of bond, which in our case is given by the bond order in the bulk.

For the case of interfaces, a similar definition is given, but this time the newly created bonds must also be considered. We have to take into account as well the addition of two different faces (A and B), each with a different number of atoms. $\Omega_\text{IBB}$ is thus given by

\begin{equation}\label{eq:bb_int}
\begin{split}
    \Omega_\text{IBB} = \frac{1}{N_\text{A}}\sum_i^{n_A} (N_{\text{b}i} - N_{\text{n}i}) \times Q_{i}
    + \frac{1}{N_\text{B}}\sum_j^{n_B} (N_{\text{b}j} - N_{\text{n}j}) \times Q_{j}
\end{split}
\end{equation}

Where $N_\text{A}$ and $N_\text{B}$ are the number of atoms at interfaces A and B and $N_{\text{n}i}$ is the number of new bonds created of type $i$.

To verify the quality of this approach, which assumes that only first neighbour interactions are relevant and that bond strength is equal for all bonds of the same type, we will perform the same analysis after relaxation using the computed bond order for each atom and each bond, which should provide a more detailed and accurate result.

For the case of surfaces, we define the change in bond strength compared to the bulk as:

\begin{equation}
\Omega_\text{SBO}=\sum_i^n \omega^{bulk}_i - \omega^{surf}_i\label{eq:bo_surf}
\end{equation}

Where $\omega^{\text{bulk}}$ is the bond order of an atom in the bulk and $\omega^{\text{surf}}$ is the bond order of an equivalent atom at the surface.

For interfaces, the change in bond strength compared to the bulk is defined as:

\begin{equation}
\Omega_\text{IBO}=\sum_i^n \omega^{\text{bulk}}_i - \omega^{int}_i\label{eq:bo_int}
\end{equation}

In this equation, $\omega^{\text{bulk}}$ is the bond order of bulk atom and $\omega^{\text{int}}$ is the bond order of an atom at the interface. The bulk calculations are done under the same strain present in the interface.

\subsubsection{Griffith Energy}

The Griffith energy is a critical parameter in understanding the mechanics of brittle fracture. It is defined as the energy required to separate an interface into two free surfaces, which equals the energy barrier for fracture propagation in brittle materials. This energy can be expressed as:

\begin{equation} \label{eq:griffith}
    \gamma_{\text{Griffith}}=\gamma_{\text{S1}}+\gamma_{\text{S2}}-E_{\text{int}}
\end{equation}

where $\gamma_{\text{Griffith}}$ is the Griffith energy, $\gamma_{\text{S}i}$ is the surface energy density of surface $i$.

\section{Study of Bagaryatskii OR interfaces with Molecular Dynamics}

\begin{figure*}[ht]
    \centering
    \includegraphics[width=0.7\textwidth]{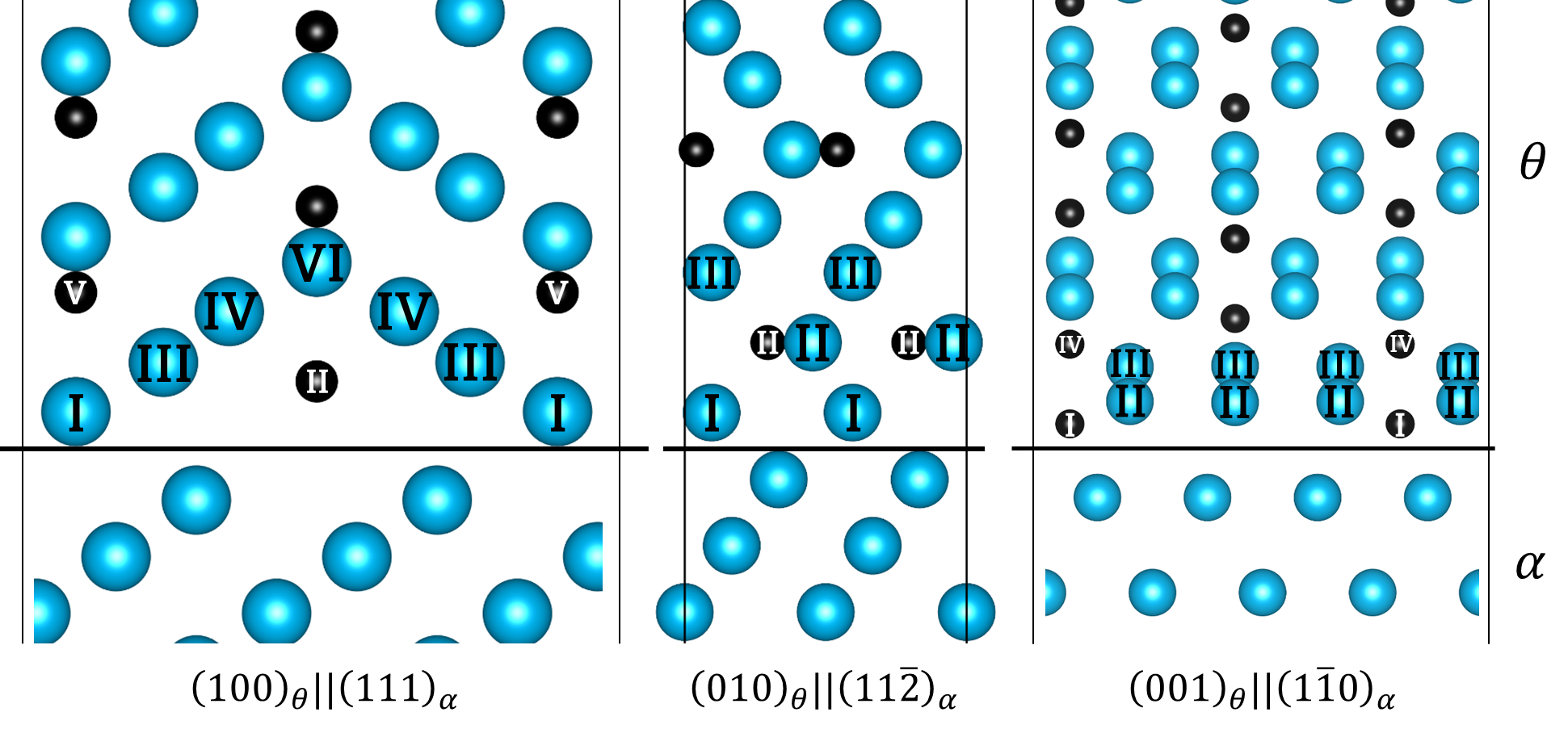}
    \caption{Left: Stoichiometric $(100)_{\theta}$ interface plane. Centre: Stoichiometric $(010)_{\theta}$ interface plane. Right: Stoichiometric $(001)_{\theta}$ interface plane. The different terminating layers are indicated by Roman numerals. Large blue spheres represent Fe atoms, and small black spheres represent C atoms. \label{fig:terminations_bag}}
\end{figure*}

In the Bagaryatskii OR, the three planes of study give rise to multiple interfaces, which are distinguished by the termination plane of the cementite. Specifically, six interfaces were identified for the $(100)_{\theta}$ plane, three for the $(010)_{\theta}$ plane, and four for the $(001)_{\theta}$ plane, as shown in Fig. \ref{fig:terminations_bag}. These are referred to as Terminations I, II, III, and so on for each interface plane. In total, this results in 13 distinct interfaces for further analysis. 

\subsection{Energetics of large (low-strain) structures}

To account for the different ways in which the cementite and ferrite layers can be arranged following the Bagaryatskii OR, we begin by constructing a significant number of interfaces. As previously mentioned, a total of 13 interfaces were built: six for the $(100)_{\theta}$ interface plane, three for the $(010)_{\theta}$ plane, and four for the $(001)_{\theta}$ plane.

In Table \ref{tab:size_bag_ORs}, we present the size and strain of each cementite lattice. It is important to note that the length in the direction perpendicular to the interface, as well as the number of atoms, are not explicitly shown, as these values depend on the terminating plane of cementite. Nevertheless, all structures are at least 50 \angstrom \space in length in this direction.

\begin{table}[ht]
    \centering
    \setlength{\tabcolsep}{3pt}
        \begin{tabular}{cccccccc}
            \cmidrule(lr){2-5}
            \multirow{2}{*}{} & \multicolumn{2}{c}{\textbf{$[010]_\theta \|[11 \overline{2}]_\alpha$}} & \multicolumn{2}{c}{\textbf{$[001]_\theta \|[1 \overline{1} 0]_\alpha$}} \\
            \cmidrule(lr){2-3} \cmidrule(lr){4-5}
            & Length (\angstrom) & Strain (\%) & Length (\angstrom) & Strain (\%) \\
            \midrule
            \midrule
            $\left( 100 \right)_{\theta}$ & 146.65 & 0.05 & 40.32 & 0.22 \\
            \bottomrule
        \end{tabular}

    \vspace{1em} 

        \begin{tabular}{cccccccc}
            \cmidrule(lr){2-5}
            \multirow{2}{*}{} & \multicolumn{2}{c}{\textbf{$[001]_\theta \|[1 \overline{1} 0]_\alpha$}} & \multicolumn{2}{c}{\textbf{$[100]_\theta \|[111]_\alpha$}} \\
            \cmidrule(lr){2-3} \cmidrule(lr){4-5}
            & Length (\angstrom) & Strain (\%) & Length (\angstrom) & Strain (\%) \\
            \midrule
            \midrule
            $\left( 010 \right)_{\theta}$ & 40.32 & 0.22 & 162.96 & 0.09 \\
            \bottomrule
        \end{tabular}

    \vspace{1em} 

        \begin{tabular}{cccccccc}
            \cmidrule(lr){2-5}
            \multirow{2}{*}{} & \multicolumn{2}{c}{\textbf{$[100]_\theta \|[111]_\alpha$}} & \multicolumn{2}{c}{\textbf{$[010]_\theta \|[11 \overline{2}]_\alpha$}} \\
            \cmidrule(lr){2-3} \cmidrule(lr){4-5}
            & Length (\angstrom) & Strain (\%) & Length (\angstrom) & Strain (\%) \\
            \midrule
            $\left( 001 \right)_{\theta}$ & 167.89 & 0.002 & 146.65 & 0.049 \\
            \bottomrule
        \end{tabular}
    \caption{Dimensions and strain of cementite for the three studied ORs along the directions of the interface planes, computed using the MEAM potential by Liyanage \cite{liyanage2014structural}. Note that the length in the direction perpendicular to the interface plane and the number of atoms vary depending on the terminating layer of cementite.}
    \label{tab:size_bag_ORs}
\end{table}

Table \ref{tab:Surface-energy-1} presents the MD surface energies of strained cementite for the three studied orientations and all identified termination planes. When comparing these values with the surface energy of unstrained cementite (shown in Table \ref{tab:Surface-energy}), it becomes evident that the effect of strain on these structures is negligible.

\begin{table}[ht]
    \centering
    \setlength{\tabcolsep}{4pt}
    \begin{tabular}{ccccccc}
            \toprule
            $\gamma_{S}\,(\text{J/m}^{2})$ & T. I & T. II & T. III & T. IV & T. V & T. VI \\
            \midrule
            \midrule
            $\left(100\right)_{\theta}$ & 2.01 & 1.97 & 2.21 & 2.09 & 2.33 & 2.18 \\
            \midrule
            $\left(010\right)_{\theta}$ & 1.80 & 1.23 & 2.25 &  &  &  \\
            \midrule
            $\left(001\right)_{\theta}$ & 2.05 & 2.11 & 2.12 & 1.68 &  &  \\
            \bottomrule
        \end{tabular}
    \caption{Surface energy $\gamma_{S}$ of strained cementite computed for the various planes and identified terminations (T) using the MEAM potential by Liyanage \cite{liyanage2014structural}.}
    \label{tab:Surface-energy-1}
\end{table}

A consequence of constructing large interfaces (with areas of 6,570 \angstrom², 24,621 \angstrom², and 5,912 \angstrom² for the three different planes) is that altering the relative position between the two crystals does not significantly affect $E_{\text{int}}$. This was verified by computing the interfacial energy for 2,500 different relative positions, totalling 32,500 calculations. As a result, we are confident that $E_{\text{int}}$ can be determined irrespective of the specific relative positioning.

Table \ref{tab:Interface-energy-of} summarises the results for $E_{\text{int}}$ across all studied planes and terminations. The three terminating atomic layers of cementite for the $(010)_{\theta}\|(11\bar{2})_{\alpha}$ interface plane show similar $E_{\text{int}}$, approximately 1 $\text{J/m}^{2}$. Three out of four terminations of the $(001)_{\theta}||(1\bar{1}0)_{\alpha}$ interface plane exhibit energies below 1 $\text{J/m}^{2}$. In contrast, all terminations of the $(100)_{\theta}||(111)_{\alpha}$ interface plane yield significantly higher $E_{\text{int}}$. Thus, we conclude that six out of the thirteen studied interfaces are energetically favourable: all three belonging to the $(010)_{\theta}\|(11\bar{2})_{\alpha}$ interface plane and terminations I, II and IV of the $(001)_{\theta}||(1\bar{1}0)_{\alpha}$ plane.

\begin{table*}[ht]
    \centering
    \begin{tabular}{ccccccc}
        \toprule
        $E_{int}\,(\text{J/m}^{2})$ & Term. I & Term. II & Term. III & Term. IV & Term. V & Term. VI \\
        \midrule
        \midrule
        $(100)_{\theta}||(111)_{\alpha}$ & 1.67 (0.69 \cite{wang2018interface}) & 1.42 & 1.36 & 1.53 & 1.41 & 1.39 \\
        \midrule
        $(010)_{\theta}||(11\bar{2})_{\alpha}$ & 1.03 (1.12 \cite{guziewski2016atomistic}, 0.11 \cite{wang2018interface}) & 1.05 (0.88 \cite{guziewski2016atomistic}) & 0.95 (2.31 \cite{guziewski2016atomistic}) &  &  &  \\
        \midrule
        $(001)_{\theta}||(1\bar{1}0)_{\alpha}$ & 0.97 (0.03 \cite{wang2018interface}) & 0.85 & 1.86 & 0.80 &  &  \\
        \bottomrule
    \end{tabular}
    \caption{Interface energy (in $\text{J/m}^{2}$) for the three studied interfacial planes and all identified terminating planes obtained using the bi-crystal method and the MEAM potential by Liyanage \cite{liyanage2014structural}. Where applicable, our results are compared with those reported in the literature. Ref. \cite{guziewski2016atomistic} shows results obtained using the MEAM potential by Liyanage. Ref. \cite{wang2018interface} provides values calculated using the EAM potential by Ruda.}
    \label{tab:Interface-energy-of}
\end{table*}

As observed in Tables \ref{tab:Surface-energy}, \ref{tab:Surface-energy-1} and \ref{tab:Interface-energy-of}, the calculated $E_{\text{int}}$ for the 13 studied systems is consistently lower than the corresponding surface energies of cementite and ferrite. This result is to be expected, given the direct relationship between surface energy and the number of broken chemical bonds. Specifically, interface energy reflects the balance between the energy required to break existing bonds and the energy associated with forming new bonds at the interface \cite{barbe2018fracture}. Thus, the formation of new bonds at the interface typically results in an interface energy that is lower than the energy of free surfaces.

Our findings can be compared with similar studies conducted by Guziewski {\it et al}. \cite{guziewski2016atomistic} in 2016 and Wang {\it et al}. \cite{wang2018interface} in 2018. Table \ref{tab:Interface-energy-of} highlights the notable differences in the interface energy values between our results and those reported in the literature.

For the $(010)_{\theta}||(11\bar{2})_{\alpha}$ interface plane, our results show that Terminations I and II align closely with those reported by Guziewski {\it et al}. \cite{guziewski2016atomistic}, who also used the MEAM potential by Liyanage \cite{liyanage2014structural}. However, there is a substantial discrepancy in the case of Termination III. Our findings, however, are more consistent with theoretical expectations, showing interface energies lower than surface energies.

Ruda \textit{et al}. find an interface energy for the stoichiometric case of 0.62 J/m² using their EAM potential. In contrast, Wang {\it et al}. \cite{wang2018interface}, using the same potential, report an interface energy of 0.11 J/m² for Termination I.

The $(001)_{\theta}||(1\bar{1}0){\alpha}$ interface is relatively underexplored in the literature. Due to this limited research, only Termination I can be directly compared with existing studies. Wang {\it et al}. \cite{wang2018interface} report a significantly lower interface energy of 0.03 J/m² for this system. Methodological differences contribute to this discrepancy: Wang {\it et al}. compute the excess energy as a function of the number of layers in the interface, and extrapolate to an infinitely thin structure to obtain $E_{\text{int}}$. However, our calculations already consider the scaling of the strain energy with the number of layers by subtracting the bulk energy of strained cementite. Furthermore, such a low $E_{\text{int}}$ would suggest a nearly perfectly coherent interface. This conclusion might be challenged by examining the lattice constants and crystal structures of the involved phases.

\subsection{The influence of strain}

Interface energy is sensitive to strain, and so it is essential to consider this variability when transitioning from large-scale structures, which are suitable for MD simulations, to smaller, more strained structures that are manageable for DFT analysis. To explore how strain affects $E_{int}$, we calculated this energy for all six interfaces as a function of strain.

\begin{figure}[ht]
    \centering
    \includegraphics[width=0.45\textwidth]{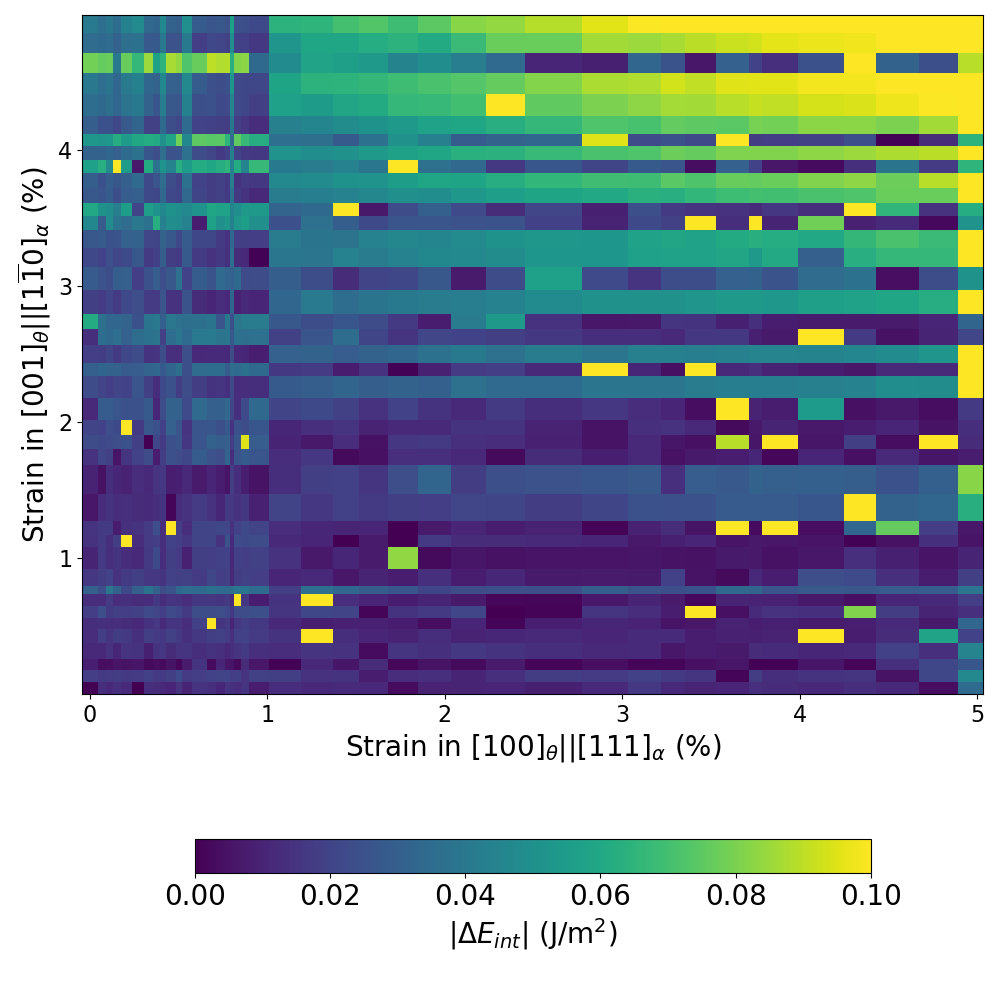}
    \caption{Influence of strain on $E_{int}$ for Termination I of the $(010)_{\theta}\|(11\bar{2})_{\alpha}$ interface. The plot shows $E_{int}$ variations with strain along the \textbf{$[100]_\theta \|[111]_\alpha$} and \textbf{$[001]_\theta \|[1 \overline{1}0]_\alpha$} axes. The colour map represents the absolute deviation $|\Delta E_{int}|$ from the minimum strain value $E_{int_{0}}$, taken as the reference. Values of $|\Delta E_{int}|$ greater than 0.10 $\text{J/m}^2$ are displayed in a single colour for better visual reference.}
    \label{fig:strain}
\end{figure}

As an example, Fig. \ref{fig:strain} shows how $E_{\text{int}}$ of the $(010)_{\theta} \parallel (\bar{1}12)_{\alpha}$, Termination I interface changes under strain along two directions, $[100]_{\theta} \parallel [111]_{\alpha}$ and $[001]_{\theta} \parallel [110]_{\alpha}$. The reference energy ($E_{\text{int},0}$) corresponds to the system at minimum strain. For strains below 3\% along $[001]_{\theta} \parallel [110]_{\alpha}$, $E_{\text{int}}$ shows minor deviations, while larger strains cause significant increases, especially along $[100]_{\theta} \parallel [111]_{\alpha}$.

Accordingly, we have obtained an estimation of how much of the discrepancy between our MD and DFT results is attributable to the additional strain, which, for less than 3\% in the directions parallel to the interface plane, should account for a maximum of 0.1 J/m², and how much is due to the difference in method.

\subsection{Energetics of small (high-strain) structures}

After identifying the most energetically favourable interfaces, we scaled down these structures to enable the DFT analysis. This scaling introduces higher strains due to the reduced number of atoms at the interface, increasing mismatch. To balance the need for a manageable number of atoms with minimal strain, we assessed strain as a function of the number of unit cells of ferrite and cementite in the previous section.
 
For the $(010)_{\theta}||(11\bar{2})_{\alpha}$ interface plane, one unit cell along the $[100]_\theta || [111]_\alpha$ axis results in a strain of 2.94\%, while seven unit cells along the $[001]_\theta || [1\overline{1}0]_\alpha$ axis yield a strain of 3.08\%. This configuration provides an interface area of 159.36 \r{A}². In contrast, for the $(001)_{\theta}||(1\bar{1}0)_{\alpha}$ interface plane, using one unit cell in each axis parallel to the interface results in strains of 4.71\% along the $[010]_\theta || [11 \overline{2}]_\alpha$ axis and 2.94\% along the $[100]_\theta || [111]_\alpha$ axis, with an interface area of 34.48 \r{A}². Although higher than desirable, trying to obtain lower strains would require a $29\times24$ ferrite supercell and a $28\times25$ cementite supercell. We therefore must assume this strain, which has been adequately assessed, if we wish to work with DFT.

Tables \ref{tab:structure_cem_bag_scale_down} and \ref{tab:interfaces_MD_bag_scale_down} detail the strain, size, and number of atoms for each of the six reduced interfaces.

\begin{table}[ht]
    \centering
    \setlength{\tabcolsep}{3pt}
        \begin{tabular}{cccccccc}
            \cmidrule(lr){2-5}
            \multirow{2}{*}{} & \multicolumn{2}{c}{\textbf{$[001]_\theta \|[1 \overline{1} 0]_\alpha$}} & \multicolumn{2}{c}{\textbf{$[100]_\theta \|[111]_\alpha$}} \\
            \cmidrule(lr){2-3} \cmidrule(lr){4-5}
            & Length (\angstrom) & Strain (\%) & Length (\angstrom) & Strain (\%) \\
            \midrule
            \midrule
            $\left( 010 \right)_{\theta}$ & 32.23 & -3.09 & 4.94 & 2.94 \\
            \bottomrule
        \end{tabular}
    \vspace{1em} 
    
        \begin{tabular}{cccccccc}
            \cmidrule(lr){2-5}
            \multirow{2}{*}{} & \multicolumn{2}{c}{\textbf{$[010]_\theta \|[11\overline{2}]_\alpha$}} & \multicolumn{2}{c}{\textbf{$[100]_\theta \|[111]_\alpha$}} \\
            \cmidrule(lr){2-3} \cmidrule(lr){4-5}
            & Length (\angstrom) & Strain (\%) & Length (\angstrom) & Strain (\%) \\
            \midrule
            \midrule
            $\left( 001 \right)_{\theta}$ & 6.98 & -4.71 & 4.94 & 2.94 \\
            \bottomrule
        \end{tabular}
    \caption{Dimensions and strains of cementite for the $\left( 010 \right)_{\theta}$ and $\left( 001 \right)_{\theta}$ planes in the case of a reduced interface, as computed using the MEAM potential by Liyanage \cite{liyanage2014structural}. Note that the length in the direction perpendicular to the interface depends on the termination layer of the cementite.}
\label{tab:structure_cem_bag_scale_down}
\end{table}

\begin{table}[ht]
    \centering
        \begin{tabular}{cccccccc}
            \cmidrule(lr){2-5}
            \multirow{2}{*}{} & \multicolumn{2}{c}{\textbf{$(010)_{\theta}||(11\bar{2})_{\alpha}$}} & \multicolumn{2}{c}{$(001)_{\theta}||(1\bar{1}0)_{\alpha}$} \\
            \cmidrule(lr){2-3} \cmidrule(lr){4-5}
            & Height (\angstrom) & $N_{atoms}$ & Height (\angstrom) & $N_{atoms}$ \\
            \midrule
            \midrule
            Term. I & 27.66 & 416 & 42.21 & 136 \\
            \midrule
            Term. II & 25.58 & 388 & 41.92 & 134 \\
            \midrule
            Term. III & 29.83 & 444 &  &  \\
            \midrule
            Term. IV &  &  & 42.39 & 138 \\
            \bottomrule
        \end{tabular}
    \caption{Characteristics of the $(010)_{\theta}||(11\bar{2})_{\alpha}$ and $(001)_{\theta}||(1\bar{1}0)_{\alpha}$ interfaces for various cementite terminations in the case of reduced interfaces, as computed using the MEAM potential by Liyanage \cite{liyanage2014structural}.}
    \label{tab:interfaces_MD_bag_scale_down}
\end{table}

As noted in Table \ref{tab:interfaces_MD_bag_scale_down}, the dimensions of the structures are smaller in the direction perpendicular to the interface plane compared to the low-strain structures, which had a minimum length of 50 \angstrom. This reduction in size is necessary to limit the number of atoms in the models.

In addition to the increased strain due to the reduced size of the structures, we expect to observe a distinct surface energy pattern. Different relative positions between ferrite and cementite may lead to varying $E_{\text{int}}$ values. Consequently, MD simulations are crucial as an intermediate step towards the construction of high-strain interfaces for subsequent DFT analysis, in order to adequately assess the influence of the relative position. The MD simulations for the high-strain structures involved computing 2500 different relative positions for each interface. The results for Term. I of the $(010)_{\theta}||(11\bar{2})_{\alpha}$ interface plane are shown in Fig. \ref{fig:gamma_termI_pbe_size} (see the Supplementary Material for the Gamma surface of the other interfaces). The Gamma surface clearly exhibits a pattern along the \textbf{$[100]_\theta ||[111]_\alpha$} axis (one unit cell), whereas the \textbf{$[001]_\theta ||[1 \overline{1} 0]_\alpha$} axis shows fewer significant oscillations in $E_{\text{int}}$.

\begin{figure}[ht]
    \centering
    \includegraphics[width=0.45\textwidth]{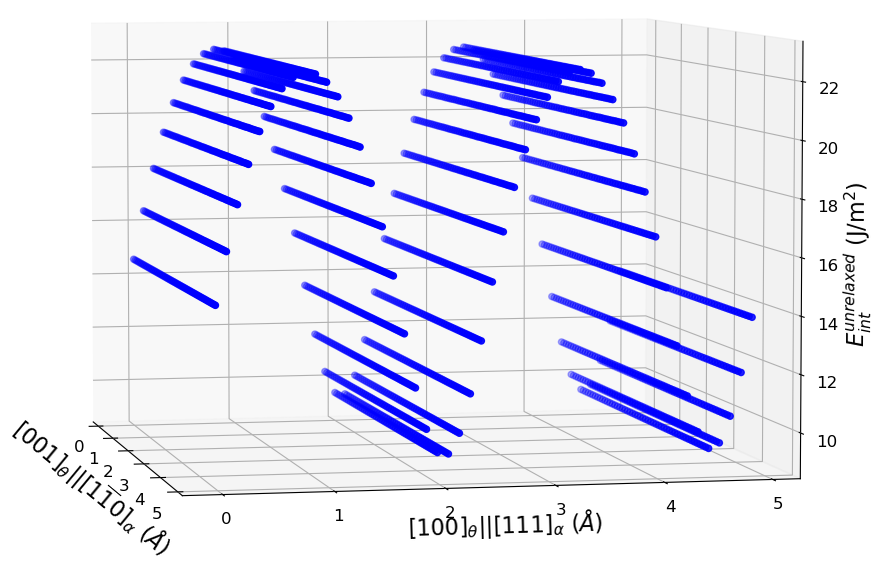}
    \caption{Gamma surface of the $(010)_{\theta}\|(11\bar{2})_{\alpha}$ interface (Term. I) in the Bagaryatskii OR. The horizontal axes represent variations in the \textbf{$[100]_\theta ||[111]_\alpha$} and the \textbf{$[001]_\theta ||[1 \overline{1} 0]_\alpha$} directions relative to a reference position, measured in \angstrom. The vertical axis shows the interface energy $E_{\text{int}}$ of the unrelaxed interface, with the lattice vector perpendicular to the interface allowed to relax.}
    \label{fig:gamma_termI_pbe_size}
\end{figure}

\begin{figure}[ht]
    \centering
    \includegraphics[width=0.30\textwidth]{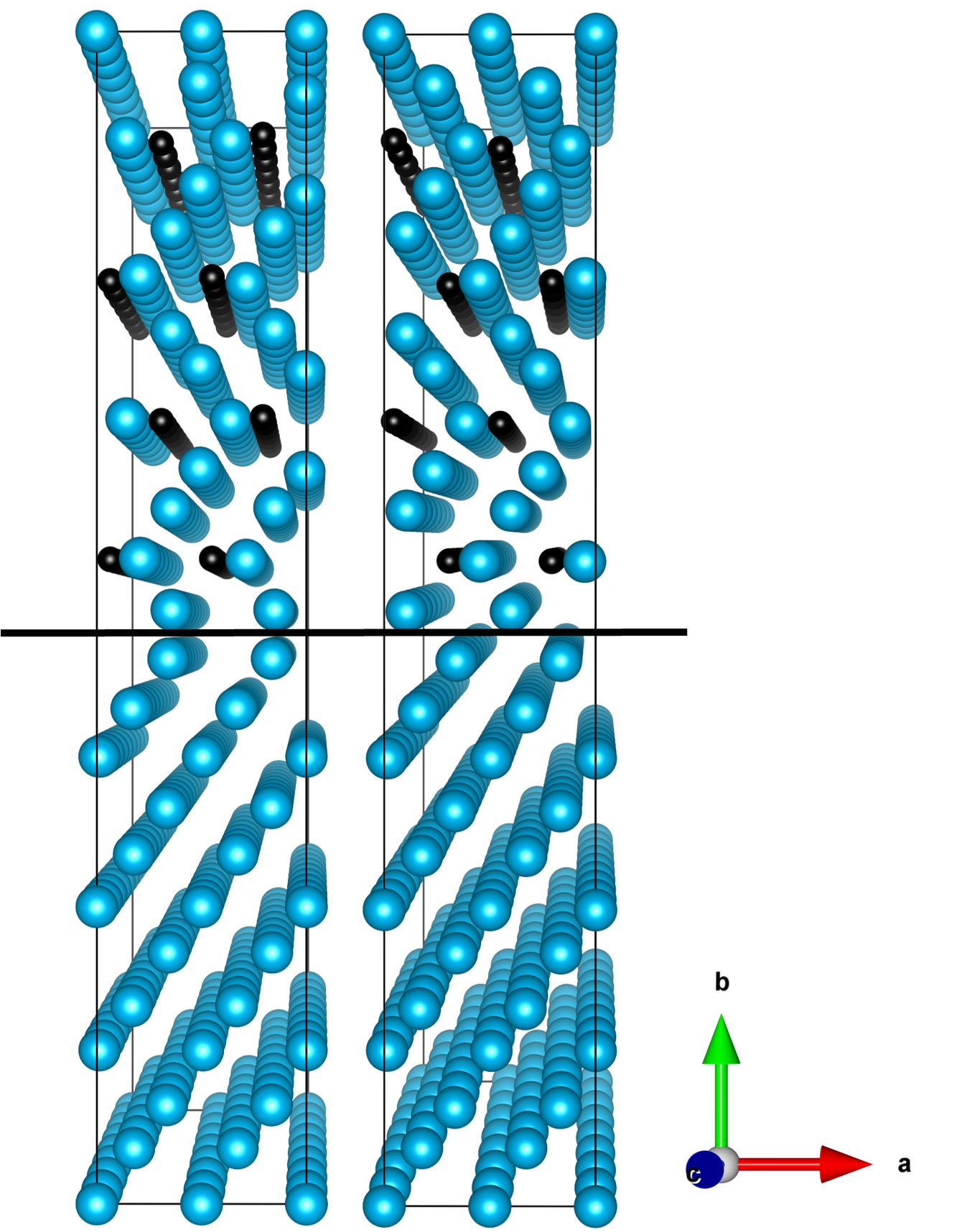}
    \caption{Ferrite and cementite relative positions that produce the highest (left) and lowest (right) $E_{\text{int}}$ for the $(010)_{\theta}||(11\bar{2})_{\alpha}$ interface (Term. I). The axis $a$ represents the $[100]_{\theta}||[111]_{\alpha}$ direction, and the axis $c$ represents the $[001]_{\theta}||[1\bar{1}0]_{\alpha}$ direction.}
    \label{fig:best_worst_termI}
\end{figure}

Fig. \ref{fig:best_worst_termI} shows the interface structures corresponding to the minimum and maximum $E_{\text{int}}$ values for Term. I of the $(010)_{\theta}||(11\bar{2})_{\alpha}$ interface. The structure minimising $E_{\text{int}}$ aligns the ferrite geometry with the cementite lattice. The maximum $E_{\text{int}}$ structure features overlapping Fe atoms, resulting in a higher interface energy.

This procedure was repeated for the remaining five terminations, yielding consistent results. We identify high-symmetry configurations that correspond to the highest and lowest interface energies: for the case of interfaces where cementite terminates with a layer that contains Fe atoms, the lowest $E_\text{int}$ occurs when Fe atoms from ferrite occupy bridge positions between cementite Fe atoms. On the other hand, the highest $E_\text{int}$ is observed when Fe atoms from ferrite and cementite are in top positions relative to each other. When the terminating layer of cementite contains carbon atoms, the lowest $E_\text{int}$ is found when Fe atoms are, on average, 1.8 \angstrom  \, from C atoms. In contrast, the highest interface energy occurs when Fe atoms are positioned closest to C atoms.

Additionally, we observed a consistent relaxation pattern in both low and high-strain interfaces. In both cases we find large deformation after relaxation in the regions where Fe atoms from ferrite and cementite sit on top of each other.

By measuring the distance between Fe atoms in ferrite, we find that, in bulk, they are found either 2.85 \angstrom or 2.47 \angstrom apart. At the interface, misfit causes some Fe atoms from ferrite to align between cementite Fe atoms at distances close to those in bulk ferrite, resulting in minimal relaxation. However, other Fe atoms from cementite sit directly on top of Fe atoms from ferrite, reducing the interatomic distance to below 2 \angstrom, which results in significant relaxation.

For the case of interfaces where the cementite terminates in C atoms, mainly terminations II and IV of interface $(001)_{\theta}||(1\bar{1}0)_{\alpha}$, relaxation increases the interatomic distance between Fe and C atoms to about 1.8 \angstrom, which is similar to the distance between Fe and C atoms of the C octahedral defect (we measure 1.83 \angstrom with the MEAM potential). Relaxation also separates Fe atoms from ferrite and cementite as to reach a distance of 2.4 or 2.8 \angstrom.

This shared behaviour suggests that strain, or at least at the studied levels, does not significantly alter the relaxation pattern at the interface. Therefore we can conclude that, even for large strains, the low-energy positions shown in the Gamma surface are correct.

By evaluating the positions that resulted in the minimum energy, we computed $E_{\text{int}}$ for the scaled-down structures, as summarised in Table \ref{tab:Eint_best_worst}.

\begin{table}[ht]
    \centering
    \begin{tabular}{ccc}
        \toprule
        $E_{int}\,(\text{J/m}^{2})$ & $(010)_{\theta}\|(11\bar{2})_{\alpha}$ & $(001)_{\theta}||(1\bar{1}0)_{\alpha}$ \\
        \midrule
        \midrule
        Term. I & 1.11 (1.03) & 0.84 (0.97) \\
        Term. II & 1.08 (1.05) & 0.53 (0.85) \\
        Term. III & 1.00 (0.95) &  \\
        Term. IV &  & 0.59 (0.80) \\
        \bottomrule
    \end{tabular}
    \caption{$E_{\text{int}}$ for the $(010)_{\theta}\|(11\bar{2})_{\alpha}$ and the $(001)_{\theta}||(1\bar{1}0)_{\alpha}$ interface planes. The starting geometries are identified from the Gamma surfaces with the minimum energy configurations. Results are obtained using MD with the MEAM potential by Liyanage \cite{liyanage2014structural}. For comparison, results obtained in the previous section with minimal strain are included.}
    \label{tab:Eint_best_worst}
\end{table}

The comparison shows that, although there are necessarily  discrepancies between the low and high strain results, these are not overly significant for the $(010)_{\theta}\|(11\bar{2})_{\alpha}$ interface. On the other hand, due to higher strains, all three $(001)_{\theta}||(1\bar{1}0)_{\alpha}$ interfaces show a larger change in energy. These results help us understand the extent to which differences between MD and DFT results are due to methodological variations versus strain effects. Furthermore, this approach allows us to avoid the impractical task of investigating the influence of relative positions with DFT.

\section{Study of Bagaryatskii OR interfaces with DFT}

Three terminations from the $(010)_{\theta}\|(11\bar{2})_{\alpha}$ interface and three from the $(001)_{\theta}||(1\bar{1}0)_{\alpha}$ interface were studied using VASP. These were selected based on their lowest energies from the thirteen originally evaluated with MD.

Table \ref{tab:structure_cem_bag_scale_down_DFT} shows the structural properties of the strained cementite crystal. The interface areas are 157.32 \r{A}² for the $(010)_{\theta}\|(11\bar{2})_{\alpha}$ and 34.03 \r{A}² for the $(001)_{\theta}||(1\bar{1}0)_{\alpha}$ interfaces. Notably, the strains are lower in DFT compared to MD for similar structures (see Table \ref{tab:structure_cem_bag_scale_down}), due to differences in lattice constants between the two methods. Table \ref{tab:interfaces_DFT_bag} provides the length of the structures perpendicular to the interface and the number of atoms for each case.

\begin{table}[ht]
    \centering
    \setlength{\tabcolsep}{3pt}
    \begin{tabular}{cccccccc}
            \cmidrule(lr){2-5}
            \multirow{2}{*}{} & \multicolumn{2}{c}{\textbf{$[001]_\theta \|[1 \overline{1} 0]_\alpha$}} & \multicolumn{2}{c}{\textbf{$[100]_\theta \|[111]_\alpha$}} \\
            \cmidrule(lr){2-3} \cmidrule(lr){4-5}
            & Length (\angstrom) & Strain (\%) & Length (\angstrom) & Strain (\%) \\
            \midrule
            \midrule
            $\left( 010 \right)_{\theta}$ & 32.04 & -1.99 & 4.91 & 2.52 \\
            \bottomrule
        \end{tabular}

    \vspace{1em} 
    \begin{tabular}{cccccccc}
            \cmidrule(lr){2-5}
            \multirow{2}{*}{} & \multicolumn{2}{c}{\textbf{$[010]_\theta \|[11\overline{2}]_\alpha$}} & \multicolumn{2}{c}{\textbf{$[100]_\theta \|[111]_\alpha$}} \\
            \cmidrule(lr){2-3} \cmidrule(lr){4-5}
            & Length (\angstrom) & Strain (\%) & Length (\angstrom) & Strain (\%) \\
            \midrule
            \midrule
            $\left( 001 \right)_{\theta}$ & 6.94 & -3.21 & 4.91 & 2.52 \\
            \bottomrule
        \end{tabular}
    \caption{Dimensions and strains of cementite for the $(010)_{\theta}||(11\bar{2})_{\alpha}$ and $(001)_{\theta}||(1\bar{1}0)_{\alpha}$ interface planes, as obtained from DFT calculations using the PBE-GGA functional. The length in the direction perpendicular to the interface varies depending on the terminating cementite layer.}
    \label{tab:structure_cem_bag_scale_down_DFT}
\end{table}

\begin{table}[ht]
    \centering
        \begin{tabular}{cccccccc}
            \cmidrule(lr){2-5}
            \multirow{2}{*}{} & \multicolumn{2}{c}{\textbf{$(010)_{\theta}||(11\bar{2})_{\alpha}$}} & \multicolumn{2}{c}{$(001)_{\theta}||(1\bar{1}0)_{\alpha}$} \\
            \cmidrule(lr){2-3} \cmidrule(lr){4-5}
            & Height (\angstrom) & $N_{atoms}$ & Height (\angstrom) & $N_{atoms}$ \\
            \midrule
            \midrule
            Term. I & 27.92 & 416 & 41.98 & 136 \\
            \midrule
            Term. II & 26.08 & 388 & 41.98 & 134 \\
            \midrule
            Term. III & 29.38 & 444 &  &  \\
            \midrule
            Term. IV &  &  & 43.21 & 138 \\
            \bottomrule
        \end{tabular}
    \caption{Structural characteristics of the $(010)_{\theta}||(11\bar{2})_{\alpha}$ and $(001)_{\theta}||(1\bar{1}0)_{\alpha}$ interfaces for the six studied cementite atomic terminations, obtained using VASP with the PBE-GGA functional.}
    \label{tab:interfaces_DFT_bag}
\end{table}

\subsection{Surface Energy Analysis}

Surface energy is a critical factor in understanding the mechanical behaviour of cementite interfaces. By analysing both strained and unstrained cases, we aim to quantify how strain impacts these surfaces at the atomic level and how much strain affects the DFT results. These findings are summarised in Table \ref{tab:Surface-energy-DFT}.

\begin{table}[ht]
    \centering
        \begin{tabular}{cccccccc}
            \cmidrule(lr){2-5}
            \multirow{2}{*}{} & \multicolumn{2}{c}{\textbf{$(010)_{\theta}$}} & \multicolumn{2}{c}{$(001)_{\theta}$} \\
            
            \cmidrule(lr){2-3} \cmidrule(lr){4-5}
            $\gamma_{S}\,(\text{J/m}^{2})$ & No strain & Strained & No strain & Strained \\
            \midrule
            \midrule
            Term. I & 2.46 & 2.42 & 2.19 & 2.22 \\
            \midrule
            Term. II & 1.81 & 1.82 & 2.54 & 2.53 \\
            \midrule
            Term. III & 2.70 & 2.68 &  &  \\
            \midrule
            Term. IV &  &  & 2.33 & 2.26 \\
            \bottomrule
        \end{tabular}
    \caption{$\gamma_{S}$ of unstrained and strained cementite for the $(010)_{\theta}$ and $(001)_{\theta}$ surface planes across various terminations and orientations. Results obtained using DFT, GGA-PBE.}
    \label{tab:Surface-energy-DFT}
\end{table}

Despite the applied strain, the surface energies of cementite in the $(010)_{\theta}$ and $(001)_{\theta}$ orientations show only a modest deviation from their original, unstrained values. This suggests that the added strain does not significantly affect the surface energy.

To explore why these strained surfaces maintain surface energies close to those of the unstrained configurations, we performed a nearest-neighbour broken bonds analysis. Using Eq. \ref{eq:bb_surf}, we calculated the surface weighted broken bonds function for each cementite surface under study.

\begin{figure}[ht]
    \centering
    \includegraphics[width=0.43\textwidth]{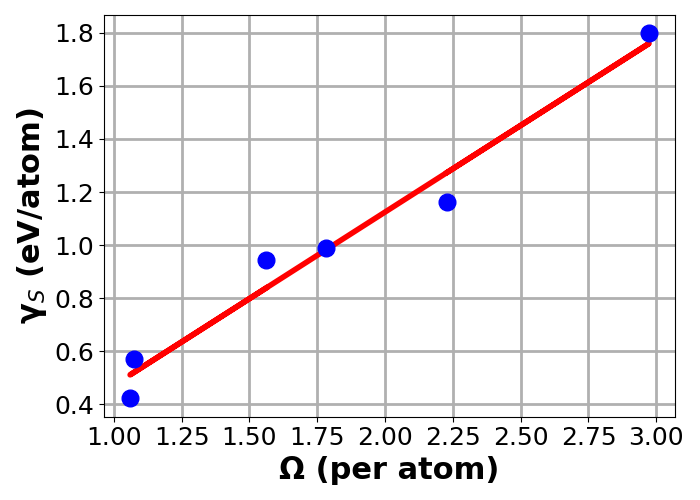}
    \caption{Correlation between the weighted number of broken bonds $\Omega$  vs. surface energy per atom for the $(010)_{\theta}$ and $(001){_\theta}$ surfaces across the three studied terminations.}
    \label{fig:bb_surface}
\end{figure}

As shown in Fig. \ref{fig:bb_surface}, the analysis reveals a linear correlation between $\Omega$ per atom and surface energy per atom. This reflects the dependence of surface energy on the number of broken bonds and the atomic density of each surface. The effects of strain, therefore, appear to be limited to influencing atomic density, while the number of broken bonds remains stable.

To account for how atomic relaxation influences bond strength and the number of broken bonds, we performed a bond order analysis on the relaxed surfaces.

\begin{figure}[ht]
    \centering
    \includegraphics[width=0.43\textwidth]{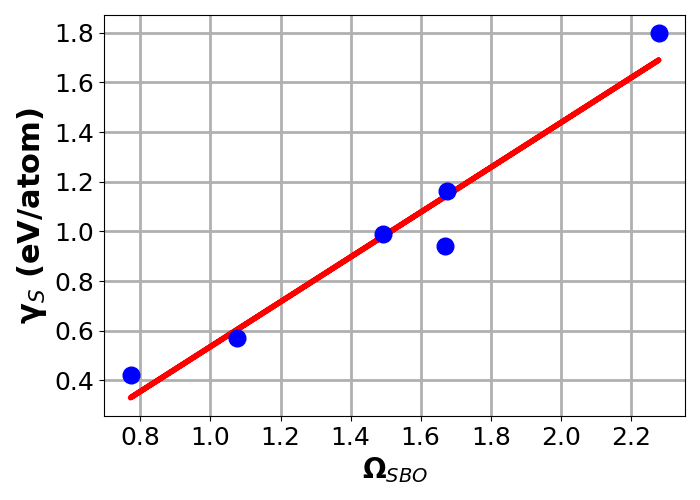}
    \caption{Correlation between the loss of bond strength $\Omega_\text{SBO}$ compared to the bulk vs. surface energy per atom for the $(010)_{\theta}$ and $(001)_{\theta}$ surfaces across the three studied terminations. $\Omega_{\text{SBO}}$ refers to the difference between the bond orders at the surface and in bulk, as defined in Eq. \ref{eq:bo_surf}.}
    \label{fig:bo_surface}
\end{figure}

As can be observed by comparing Figs. \ref{fig:bb_surface} and \ref{fig:bo_surface}, relaxing the surfaces does not significantly alter the overall broken bond structure. Therefore, the broken bond method can be reliably used as a preliminary tool to study surfaces without requiring extensive computational resources.

\subsection{Interface Energy Analysis}

Table \ref{tab:Eint_DFT} presents the interface energies $E_{\text{int}}$ for the six studied interfaces calculated using VASP with the GGA-PBE functional with the methodologies detailed in sections \ref{sec:Molecular_Dynamics_Methods} and \ref{sec:Eint}.

\begin{table}[ht]
    \centering
    \begin{tabular}{ccc}
        \toprule
        $E_{int}\,(\text{J/m}^{2})$ & $(010)_{\theta}\|(11\bar{2})_{\alpha}$ & $(001)_{\theta}||(1\bar{1}0)_{\alpha}$ \\
        \midrule
        \midrule
        Term. I & 1.17 (0.45 \cite{zhang2015structural}) & 0.65 \\
        Term. II & 1.23 & 0.58 \\
        Term. III & 1.09 &  \\
        Term. IV &  & 0.67\\
        \bottomrule
    \end{tabular}
    \caption{$E_{\text{int}}$ of the $(010)_{\theta}\|(11\bar{2})_{\alpha}$ and the $(001)_{\theta}||(1\bar{1}0)_{\alpha}$ interface planes for the six studied atomic terminations of cementite. Results obtained with VASP using the GGA-PBE scheme. When possible, our results are compared with those found in the literature. Ref. \cite{zhang2015structural} obtained using GGA-PBE.}
    \label{tab:Eint_DFT}
\end{table}

Our analysis reveals that the most energetically favourable interfaces are consistently found to be the three studied terminations of the $(001)_{\theta}||(1\bar{1}0)_{\alpha}$ interface plane. This observation is consistent with the MD results reported in Tables \ref{tab:Interface-energy-of} and \ref{tab:interfaces_MD_bag_scale_down}, reinforcing the idea that this interface plane generally exhibits lower interface energies.

The interface energies obtained from DFT for the $(010)_{\theta}\|(11\bar{2})_{\alpha}$ interface are consistently higher than those from MD simulations. Although some of this discrepancy can be attributed to the effects of strain, which were estimated to contribute up to 0.1 $\text{J/m}^2$, a significant portion of the difference is likely due to the superior accuracy of the DFT method compared to MD.
        
For the $(001)_{\theta}||(1\bar{1}0)_{\alpha}$ interface plane, DFT results show different $E_{int}$ values compared to MD results that range between 0.05 and 0.19 J/m$^2$.

Interfaces without carbon in the cementite termination layer (Fig. \ref{fig:terminations_bag}, Terminations III and II of the $(010)_{\theta}||(11\bar{2})_{\alpha}$ and the $(001)_{\theta}||(1\bar{1}0)_{\alpha}$ interface planes respectively) have the lowest $E_{\text{int}}$, while those with carbon (Terminations II and IV of the $(010)_{\theta}||(11\bar{2})_{\alpha}$ and the $(001)_{\theta}||(1\bar{1}0)_{\alpha}$ interface planes) exhibit higher values, highlighting carbon's destabilizing role, contrary to what previous studies had found \cite{guziewski2016atomistic}.

We use our nearest neighbour bond analysis to find a correlation between the weighted contribution of newly created and broken bonds as described in Eq. \ref{eq:bb_int} and interface energy per atom. To account for how atomic relaxation influences bond strength and the number of broken bonds, we performed a bond order analysis on the relaxed surfaces.

\begin{figure}[ht]
    \centering
    \includegraphics[width=0.43\textwidth]{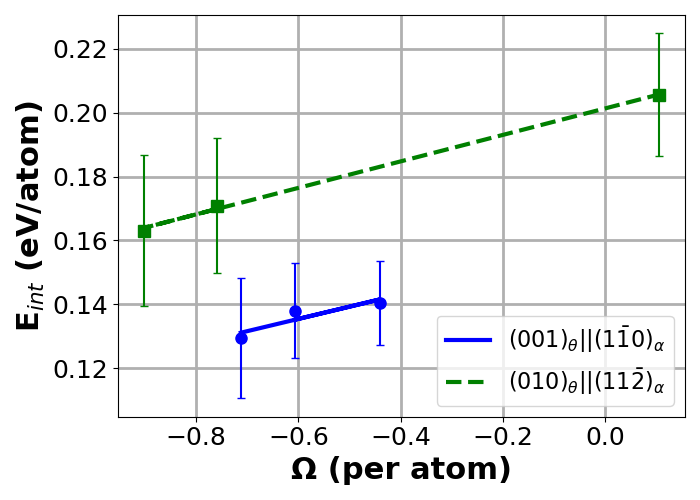}
    \caption{Correlation between the weighted number of broken and new bonds $\Omega$ per atom  vs. interface energy per atom for the $(010)_{\theta}\|(11\bar{2})_{\alpha}$ and $(001)_{\theta}||(1\bar{1}0)_{\alpha}$ interfaces across the three studied terminations.}
    \label{fig:bb_int}
\end{figure}

\begin{figure}[ht]
    \centering
    \includegraphics[width=0.43\textwidth]{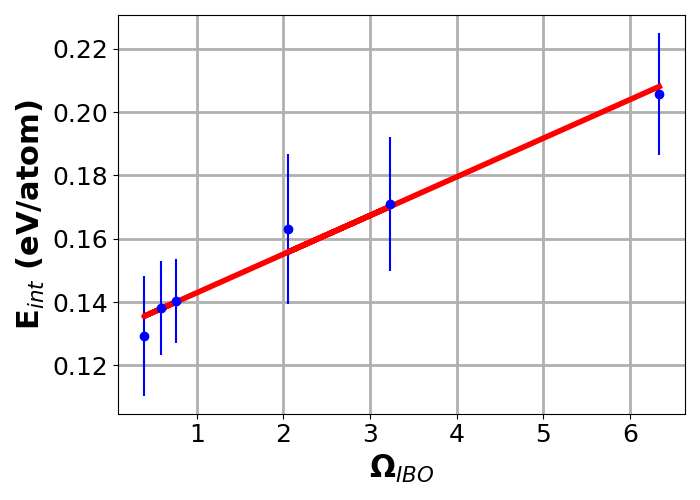}
    \caption{Correlation between the change in bond strength compared to the bulk $\Omega_\text{IBO}$ vs. interface energy per atom for the $(010)_{\theta}\|(11\bar{2})_{\alpha}$ and $(001)_{\theta}||(1\bar{1}0)_{\alpha}$ interfaces for the the three studied terminations. $\Omega_{\text{IBO}}$ refers to the difference between the bond orders at the interface and in bulk, as defined in Eq. \ref{eq:bo_int}.
    }
    \label{fig:bo_int}
\end{figure}

Figure \ref{fig:bb_int}  shows two different linear correlations between the weighted number of broken and newly formed bonds ($\Omega$) and the interface energy per atom ($E_{\text{int}}$) for the $(010)_{\theta}||(11\bar{2})_{\alpha}$ and the $(001)_{\theta}||(1\bar{1}0)_{\alpha}$ interfaces. $\Omega$ considers the atomic positions prior to relaxation to enable predictions without the need of extensive computational resources. The difference in trends may then reflect the influence of strain relaxation perpendicular to the interface, which alters the interfacial distance and hence affects the bond distribution and energy stabilization. In the case of the $(001)_{\theta}||(1\bar{1}0)_{\alpha}$ configuration, a higher strain results in a higher relaxation in the perpendicular direction, thus explaining the different correlation. This does not occur for the bond order analysis, shown in Fig. \ref{fig:bo_int}, which is computed using the relaxed structures and all individual bond orders.
The results reinforce the importance of accurately describing bond reorganization during interface formation.

The bond order analysis differs from the usual implementation of the generalised nearest-neighbour broken-bond analysis \cite{sonderegger2009generalized, gunda2023first}, which considers only broken bonds. The latter approach would approximate interfaces as two separate surfaces that do not interact with each other, whereas our approach considers the stabilising factor of new bond creation. Using our method and taking the $(010)_{\theta}\|(11\bar{2})_{\alpha}$ linear regression equation, for a similarly strained $(100)_{\theta}||(111)_{\alpha}$ interface, we obtain a value of $E_{\text{int}}=1.58 \,\text{J/m}^2$. Considering just broken bonds, this value changes to 1.76 $\,\text{J/m}^2$. On the other hand, MD yields an energy of 1.62 J/m$^2$ for the strained structure and of 1.67 J/m$^2$ for the case with low strain (see Table \ref{tab:Interface-energy-of}), revealing the predicting capabilities of the method that considers both newly created and broken bonds.

\begin{figure*}
    \centering
    \includegraphics[width=0.85\linewidth]{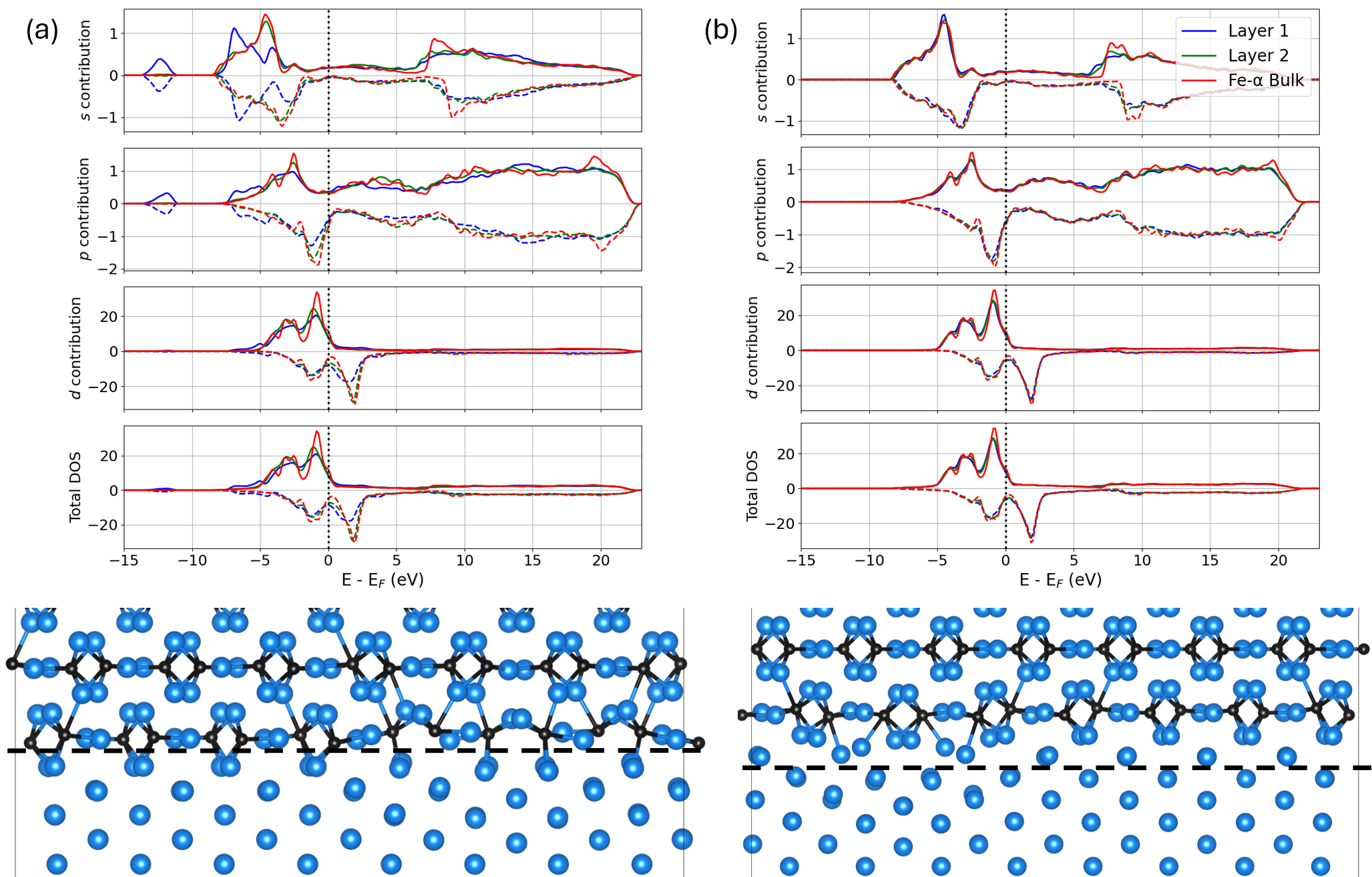}
    \caption{Projected Density of States (PDOS) for the $s$, $p$, and $d$ orbitals, as well as the total PDOS, in the two outermost atomic layers of ferrite, compared with the bulk state. Results are shown for two representative interfaces: $(010)_{\theta}\|(11\bar{2})_{\alpha}$ terminations II (a), where interfacial carbon induces Fe–C hybridization, and III (b), featuring a pure Fe cementite termination. The lower panels show the relaxed atomic structures. Red lines correspond to bulk ferrite, blue to the outermost interface layer, green to the second outermost layer, and the vertical dashed line marks the Fermi energy.}
    \label{fig:dos}
\end{figure*}

Finally, we examine the relationship between the electronic structure of the interfaces and their energy. To this end, we first compute the electronic Projected Density of States (PDOS) for the ferrite atomic layers as a function of their position relative to the interface. The most representative results are shown in Fig. \ref{fig:dos} for the $(010)_{\theta}\|(11\bar{2})_{\alpha}$ interface, terminations II and III (PDOS for all studied interfaces can be found in the supplementary material).

As observed, when the terminating layer of cementite contains carbon atoms, the PDOS of the two outermost ferrite layers is significantly altered compared to their bulk state. For example, in the $(010)_{\theta}\|(11\bar{2})_{\alpha}$ interface, termination II, shown in Fig. \ref{fig:dos} (a), substantial changes appear in the $s$ and $p$ orbital contributions, which are associated with bonding to C atoms. This suggests that the local bonding environment is more strongly modified in these cases, leading to an increase in $E_\text{int}$.

In contrast, the interface $(010)_{\theta}\|(11\bar{2})_{\alpha}$, termination III, where cementite terminates in two Fe layers, exhibits minimal deviations in PDOS. The small differences compared to the bulk ferrite PDOS indicate that, after relaxation, bonding is partially restored, supporting the trends presented in Figs. \ref{fig:bb_int} and \ref{fig:bo_int}.

\begin{figure}
    \centering
    \includegraphics[width=1\linewidth]{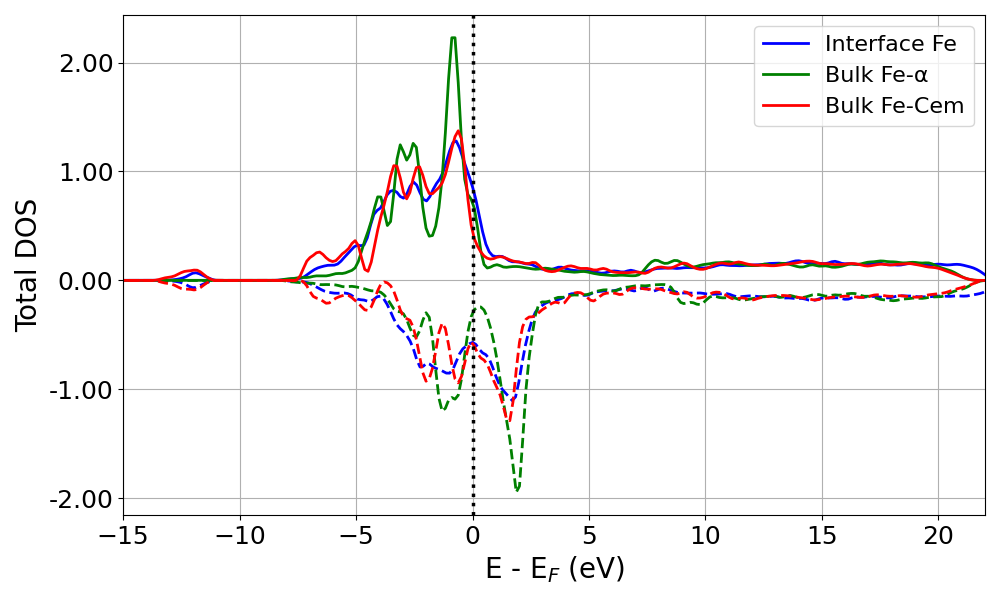}
    \caption{Projected Density of States (PDOS) for a Fe atom near the $(010)_{\theta}\|(11\bar{2})_{\alpha}$ interface, termination II from ferrite (blue line) compared to the bulk PDOS of a Fe I atom in cementite (red line) and a Fe atom in ferrite (green line).}
    \label{fig:dos_Fe_C}
\end{figure}

Figure \ref{fig:dos_Fe_C} provides a comparison of the PDOS for a representative interfacial Fe atom from ferrite (blue line) and we compare it with the bulk state of a Fe atom from cementite (red line) and a bulk Fe atom from ferrite (green line). It is clear that the Fe atoms from the terminating ferrite layer in interface $(010)_{\theta}\|(11\bar{2})_{\alpha}$, termination II, adopt a PDOS profile similar to that of Fe atoms in cementite. In contrast, we find that the projected PDOS of the C atoms (shown in Fig. 9 of the supplementary material) remains nearly identical to that of bulk cementite, indicating that their local electronic structure is mostly unaffected by the presence of the interface. This asymmetric response suggests that destabilization arises primarily from perturbations of the local electronic structure of the Fe-$\alpha$ atoms at the interface.

\begin{figure}[ht]
    \centering
    \includegraphics[width=0.5\textwidth]{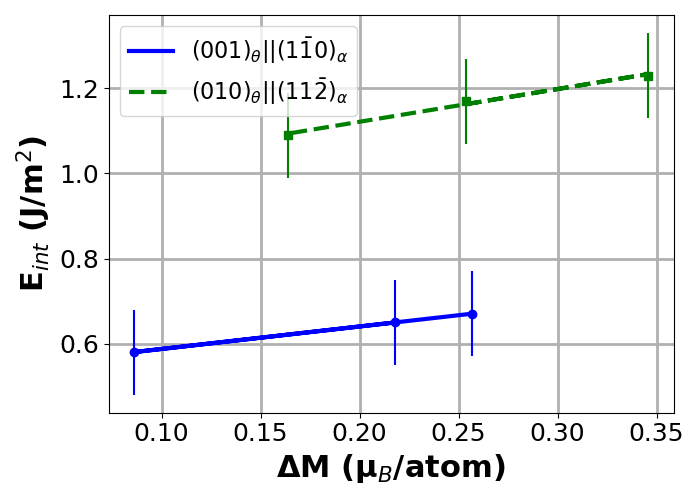}
    \caption{Correlation between the magnetic moment variation of the terminating layer of ferrite vs. interface energy for the $(010)_{\theta}\|(11\bar{2})_{\alpha}$ and $(001)_{\theta}||(1\bar{1}0)_{\alpha}$ interfaces across the six studied terminations.}
    \label{fig:magmom_int}
\end{figure}

In addition, we study the variation in the atomic magnetic moment $\Delta M$ in the terminating ferrite layer, which is directly related to the PDOS and, in turn, to the bond structure and coordination, and is thus linked to the interface energy of each system \cite{blonski2007structural, wachowicz2010cohesive}. The change in the PDOS is most relevant in the $d$-band of Fig. \ref{fig:dos}(a)  (please note the different scale for the $y$ axis used for $p$ and $d$ orbitals). In particular, we observe a decrease in the majority-spin (spin-up) $d$-states and a corresponding increase in the minority-spin (spin-down) $d$-states near the Fermi level. The $p$-orbital contributions, in contrast, remain mostly unchanged near $E_\text{F}$. Although these differences are subtle and not visually dominant in the PDOS plots, they reflect a redistribution of electronic states between spin channels that leads to a reduction in spin polarization. It is therefore a sum of various minor changes in the spin-up and spin-down PDOS which leads to the observed change in the magnetic moment of interfacial Fe atoms, in agreement with the trend shown in Fig. \ref{fig:magmom_int}.

By integrating the spin-up and spin-down curves and calculating their difference, we find that $\Delta M$ increases for interfaces where the cementite terminating layer contains carbon atoms. Moreover, across all interface orientations, we obtain a linear correlation similar to the one obtained for the relationship between the weighted number of broken bonds $\Omega$ and $E_{\text{int}}$ (see Fig. \ref{fig:bb_int}). All these findings suggest a stronger perturbation of the electronic structure at the  C-terminated interfaces and are consistent with  their increased $E_\text{int}$ with respect to the Fe-terminated ones.

\subsubsection{Griffith Energy}

Table \ref{tab:griffith_energy} shows the interface energy, surface energies of cementite and ferrite and Griffith energy as calculated with Eq. \ref{eq:griffith} for all of the studied systems using DFT.

\begin{table}[ht]
\centering
\setlength{\tabcolsep}{4pt}
\begin{tabular}{lcccc}
    \toprule
    Interface & $E_\text{int}$ & $\gamma_\text{Cem}$ & $\gamma_\text{Fe}$ & $\gamma_\text{Griffith}$ \\
    \midrule
    \midrule
    $(010)_\theta||(11\bar{2})_\alpha$ Term. I & 1.17 & 2.42 & 2.58 & 3.84\\
    $(010)_\theta||(11\bar{2})_\alpha$ Term. II & 1.23 & 1.82 & 2.58 & 3.17 \\
    $(010)_\theta||(11\bar{2})_\alpha$ Term. III & 1.09 & 2.68 & 2.58 & 4.17\\
    $(001)_\theta||(1\bar{1}0)_\alpha$ Term. I & 0.65 & 2.22 & 2.44 & 4.00 \\
    $(001)_\theta||(1\bar{1}0)_\alpha$ Term. II & 0.58 & 2.53 & 2.44 & 4.39 \\
    $(001)_\theta||(1\bar{1}0)_\alpha$ Term. IV & 0.67 & 2.26 & 2.44 & 4.02\\
    \bottomrule
\end{tabular}
\caption{Calculated interface energy, surface energies of cementite and ferrite, and Griffith energy for the studied systems using DFT, in units of J/m$^2$.}
\label{tab:griffith_energy}
\end{table}

For the interface plane $(010)_{\theta}\|(11\bar{2})_{\alpha}$, we find that the Griffith energy is the highest for termination III. This corresponds to the case where carbon is not present in the first two surface atomic layers. The next highest energy, termination I, corresponds to a Fe terminating layer, with carbon being located in the second layer. Finally, the lowest Griffith energy is computed for termination II, which presents carbon in the terminating atomic layer.

For the  $(001)_{\theta}||(1\bar{1}0)_{\alpha}$ interface, we find the same situation. Terminations I and IV present carbon atoms at the terminating layer, thus their Griffith energies are the lowest. On the other hand, termination II has the highest energy, corresponding to the case with no carbon atoms in the first two terminating layers. We can thus conclude that carbon atoms play a key role in interface stability against fractures, which can be linked to a higher number of broken bonds at the interface.

\section{Conclusions}

This study presents a systematic multiscale approach that addresses several open questions in the literature concerning the Bagaryatskii Orientation Relationship in ferrite–cementite interfaces. Our method bridges the gap between large-scale Classical Molecular Dynamics (MD) simulations and high-accuracy Density Functional Theory (DFT) calculations, resolving discrepancies in previously reported interface energies and enabling a consistent treatment of non-stoichiometric cementite terminations.

Among the configurations studied, the $(010)_{\theta}||(11\bar{2})_{\alpha}$ and $(001)_{\theta}||(1\bar{1}0)_{\alpha}$ interfaces emerged as the most stable. Trends observed in MD simulations were corroborated by DFT calculations, which also provided access to new electronic and magnetic properties that are difficult to assess using empirical methods alone. These results offer valuable insight into the atomic-scale mechanisms governing interface stability.

Importantly, we find that interfaces terminated by carbon atoms exhibit higher interface energies and lower Griffith energies, indicating reduced resistance to fracture. This behaviour contrasts with earlier studies suggesting a stabilizing role for interfacial carbon. Our analysis reveals that, while carbon atoms retain a cementite bulk-like electronic environment at the interface, Fe atoms experience significant perturbations in their electronic structure and magnetic moment due to Fe–C bonding. This asymmetry underlies the destabilizing influence of interfacial carbon.

Additionally, we identify robust correlations between interface energy, local magnetic moment variation in ferrite, and a bond-based metric quantifying broken and newly formed bonds. These relationships support a physically grounded, predictive framework for evaluating interface stability across similar systems.

Overall, this work not only advances our understanding of ferrite–cementite interfaces but also demonstrates the potential of combined DFT/MD strategies and bond analysis tools for designing steels with improved mechanical performance under demanding conditions. Future extensions may incorporate the role of solutes, point defects, or hydrogen, as well as explore the use of machine-learning-based potentials to enable efficient, accurate modelling of even more complex interfacial phenomena.

\section*{CRediT authorship contribution statement}

\textbf{Pablo Canca:} Conceptualization, Formal analysis, Investigation, Visualization, Writing - original draft, Writing - review and editing. \textbf{Chu Chun Fu:} Conceptualization, Methodology, Supervision, Writing - review and editing. \textbf{Christophe Ortiz:} Conceptualization, Methodology, Funding acquisition, Project administration, Resources, Supervision, Writing - review and editing. \textbf{Blanca Biel:} Conceptualization, Methodology, Funding acquisition, Project administration, Resources, Supervision, Writing - review and editing.

\section*{Declaration of competing interest}
The authors declare that they have no known competing financial interests or personal relationships that could have appeared to influence the work reported in this paper.

\section*{Declaration of Generative AI and AI-assisted technologies in the writing process}
During the preparation of this work the authors used ChatGPT in order to assist in the correction and editing of the English manuscript. After using this tool, the authors reviewed and edited the content as needed and take full responsibility for the content of the publication.

\section*{Acknowledgements}
This work has been supported by the European Union’s FEDER program, IFMIF-DONES Junta de Andalucia’s program at the Universidad de Granada (SE21 IFMIF-DONES FEDER).
The authors thankfully acknowledge the technical expertise and assistance provided by the Spanish Supercomputing Network (Red Española de Supercomputación), as well as the computer resources used: the LaPalma Supercomputer, located at the Instituto de Astrofísica de Canarias.
Part of the numerical calculations were performed using Marconi supercomputer facility at CINECA, Bologna, Italy, supported by EUROfusion.
Authors would like to thank the computing time provided by the Servicio de Supercomputación de la Universidad de Granada (\url{https://supercomputacion.ugr.es}). We also want to thank Pedro Delgado for our fruitful discussions.

\bibliographystyle{elsarticle-num} 
\bibliography{references.bib}

\end{document}